\documentclass[preprint,12pt]{elsarticle}
\usepackage{amssymb}
\usepackage[utf8]{inputenc}
\usepackage{amsmath}
\usepackage{amsfonts}
\usepackage{amssymb}
\usepackage{multirow}
\usepackage{graphicx}
\usepackage{hyperref}
\usepackage{subfig}
\usepackage[toc,page]{appendix}
\usepackage{multirow}
\usepackage{natbib}
\usepackage{hyperref}

\journal{Physics of the Dark Universe}

\begin{document}

\begin{frontmatter}

\title{Cosmological constraints on parametrized interacting dark energy}

\author{R. von Marttens}
\address{N\'ucleo COSMO-UFES \& Departamento de F\'isica,  Universidade Federal do Esp\'irito Santo (UFES).
 Av. Fernando Ferrari s/n CEP 29.075-910, Vit\'oria, ES, Brazil.}
\address{Departamento de Astronomia, Observat\'orio Nacional, 20921-400, Rio de Janeiro, RJ, Brasil.}
\address{D\'epartement de Physique Th\'eorique and Center for Astroparticle Physics, Universit\'e de Gen\`eve, 24 quai Ernest Ansermet, CH-1211
Geneva, Switzerland}
\author{L. Casarini}
\address{International Institute of Physics (IIP), Universidade Federal do Rio Grande do Norte (UFRN) CP 1613, 59078-970 Natal-RN, Brazil.}
\address{Institute of Theoretical Astrophysics, University of Oslo, 0315 Oslo, Norway.}
\author{D.F. Mota}
\address{Institute of Theoretical Astrophysics, University of Oslo, 0315 Oslo, Norway.}
\author{W. Zimdahl}
\address{N\'ucleo COSMO-UFES \& Departamento de F\'isica,  Universidade Federal do Esp\'irito Santo (UFES).
 Av. Fernando Ferrari s/n CEP 29.075-910, Vit\'oria, ES, Brazil.}
 
\begin{abstract}
We reconsider the dynamics of the Universe in the presence of interactions in the cosmological dark sector. A class of interacting models is introduced via a real function $f\left(r\right)$ of the ratio $r$ between the energy densities of the (pressureless) cold dark matter (CDM) and dark energy (DE).
The subclass of models for which the ratio $r$ depends only on the scale factor is shown to be equivalent to unified
models of the dark sector, i.e. models for which the CDM and DE components can be combined in order to form a unified
dark fluid. For specific choices of the function $f\left(r\right)$ we recover several models already studied in the literature.
We analyse various special cases of this type of interacting models using a suitably modified version of the CLASS
code combined with MontePython in order to constrain the parameter space with the data from supernova of type SNe Ia
(JLA), the Hubble constant $H_{0}$, cosmic chronometers (CC), baryon acoustic oscilations (BAO) and data from the Planck satellite (Planck TT).
Our analysis shows that even if data from the late Universe ($H_{0}$, SNe Ia and CC) indicate an interaction in
the dark sector, the data related to the early Universe (BAO and Planck TT) constrain this interaction substantially, in particular for cases in which the background dynamics is strongly affected.
\end{abstract}


\end{frontmatter}

\section{Introduction}
\label{intro}

One of the most intriguing challenges of current cosmology is the nature of the dark sector of the Universe. According the most recent observations \cite{Abbott:2017smn,Ade:2015xua,Ade:2015rim,Betoule:2014frx,Kowalski:2008ez,Hicken:2009dk,Amanullah:2010vv,Suzuki:2011hu,Hinshaw:2012aka}, we live in a spatially flat Universe and this dark sector contributes with approximately  95\% to the cosmic substratum today. The rest of the material content of the Universe is composed by a negligible part of radiation and the remaining 4-5\% by baryonic matter, the kind of matter that composes systems that interact with electromagnetic radiation, and therefore can be observed directly (like ourselves!).

Each of the components of the dark sector plays an important role in the dynamics of the Universe. The dark matter, which corresponds to 25\% of the matter content of the Universe, is an exotic pressureless matter which was proposed to explain the observations of the velocity of galaxy clusters \cite{Zwicky:1933gu}. Years later, the existence of the CDM was corroborated with the studies of the rotation curves of spiral galaxies \cite{Sofue:2000jx}, which indicated that there was more mass in galaxies than could be observed through their luminosity. Moreover, the analysis of x-ray emission  by galaxy clusters and gravitational lensing also indicates the presence of this exotic matter. In the context of structure formation, CDM seems to play a very important role, potentializing the growth of baryonic structures after decoupling, until they reach the non-linear regime that is currently observed ($\delta_{b}>1$).

The dark energy, which is responsible for the remaining 70\% of the cosmic substratum was proposed to explain the current phase of accelerated expansion of the Universe \cite{Riess:1998cb,Perlmutter:1998np}. Within the cosmological standard description, the DE component can be identified with the cosmological constant $\Lambda$, which \textit{a priori} has a geometric nature in the context of the general theory of relativity. Such identification is analogous to a fluid model with a vacuum equation of state (EoS) $w=-1$ and constant energy density. As previously mentioned, this description of dark energy seems to successfully satisfy the most recent observational data, however it is in deep disagreement with the theoretical prediction for vacuum energy that comes from quantum field theory \cite{Weinberg:1988cp}.

Along with the general theory of relativity (GR), the inflationary paradigm and the Big Bang nucleosynthesis (BBN), this material description composes the so-called $\Lambda$CDM model.
Instead of the vacuum description, it is also common to consider a dynamical description for the DE component through a different EoS, for example, a constant EoS parameter
$w\neq -1$  or some time dependent EoS parameter \cite{Chevallier:2000qy,Linder:2002et}. Alternatively, several alternatives to describe the DE component are proposed in the literature, among others, the dynamical approach through a scalar field \cite{Caldwell:1997ii,ArmendarizPicon:2000ah} and modified theories of gravity \cite{Clifton:2011jh,Joyce:2014kja} have received much attention.

In this work, we focus on the study of cosmological models in which, unlike in the standard cosmological description, CDM and DE are not independent components, but there is a non-gravitational interaction that results in an energy exchange between them. An important feature of this class of models is that such interaction implies the existence of DE perturbations even in the case where $w=-1$. This type of models has been extensively studied in the literature \cite{Billyard:2000bh,Amendola:1999er,Zimdahl:2001ar,Chimento:2003iea,He:2008tn,Valiviita:2008iv,Clemson:2011an,Li:2014eha,Skordis:2005xk,Akrami:2013ffa,Barrow:2002zh,Faraoni:2014vra,Marra:2015iwa,Odderskov:2015fba,Wang:2016lxa,Marttens:2016cba,Arevalo:2016epc,Kumar:2017dnp,Gonzalez:2018rop} as a simple and viable alternative to the standard cosmological model, and there are arguments that indicate that it is not correct to ignore this interaction \cite{Das:2005yj,Abdalla:2014cla} or to ignore the DE perturbations in dynamic DE models \cite{Hwang:2009zj}. Recent studies indicate some remarkable observational aspects of these interacting models \cite{Santos:2017bqm,Mifsud:2017fsy,Gomez-Valent:2017idt}.

In general, the motivation for these models is phenomenological (although some cases may be based on a more fundamental argument \cite{Sola:2011qr}) and each model is seen completely independently of the others. Furthermore, most of the interacting models proposed in the literature are such that the interaction term is linear \cite{Costa:2016tpb}, i.e., it depends only linearly on one of the energy densities of the dark sector components. Here, we propose a more general description in which the interaction term is a real function of the energy densities of CDM and DE. This approach allows us to relate several models of interaction through a function of the ratio between the energy densities of CDM and DE. For specific cases we find analytical solutions, some of them already present in the literature.

This paper is organized as follows: In Section \ref{background} we introduce a background description of the interaction between CDM and DE via a real function of the ratio between CDM and DE energy densities. We recover several cases already studied in the literature and we demonstrate the equivalence of a class of interacting models with unified models, i.e., models that can be described as a single perfect conservative fluid.
The well-established linear perturbation theory for  interacting perfect fluids is recalled in Section \ref{perturbations}. In Section \ref{specific} we present some specific cases of interacting models obtained through the proposed generalization. In Section \ref{analysis} we perform a statistical analysis for each model using the observational data from SNe Ia (JLA) \cite{Betoule:2014frx}, local measures of $H_{0}$ \cite{Riess:2011yx}, BAO \cite{Beutler:2011hx,Ross:2014qpa,Anderson:2013zyy,Alam:2016hwk,Kazin:2014qga} and the CMB temperature anisotropy spectrum. Finally, Section \ref{conclusions} summarizes our results.

\section{Background dynamics of interacting models}
\label{background}

\subsection{General equations}

At the background level the Universe is considered to be homogeneous, isotropic and spatially flat  and  describable  by the FLRW metric
\begin{equation}
ds^{2}=dt^{2}-a^{2}\left(t\right)\left[dr^{2}+r^{2}\left(d\theta^{2}+\sin^{2}\theta d\phi\right)\right],
\label{flrw}
\end{equation}
where $a$ is the scale factor. The expansion dynamics obeys Friedmann's equation
\begin{eqnarray}
H^{2}=\frac{8\pi G}{3}\rho,\label{friedmann}
\end{eqnarray}
and
\begin{eqnarray}
\dot{H}=-4\pi G\left(\rho+p\right),
\end{eqnarray}
where $H\equiv\dot{a}/a$ is the Hubble rate, $\rho$ and $p$ are, respectively, the total energy density and the pressure of the material content of the Universe . Considering a GR context, the total cosmic fluid must be conservative,
\begin{equation}
\dot{\rho}+3H\left(\rho+p\right)=0.
\label{universe}
\end{equation}

We assume that the material content of the universe is composed of four components: radiation, baryons, CDM and DE, all of them described by ideal fluids with EoS $p_{i}=w_{i}\,\rho_{i}$. The radiation component will be denoted by a subindex $r$ and it is characterized by a state parameter $w_{r}=1/3$. Baryons  are a pressureless component, and will be denoted by a subindex $b$ ($w_{b}=0$). The CDM component is also pressureless, and will be denoted by a subindex $c$ ($w_{c}=0$). Lastly, the DE component will be denoted by a subindex $x$, it is characterized by a constant EoS parameter $w_{x}=-1$, which can be associated to a cosmological constant. The total energy density $\rho$ and total pressure $p$ are defined as the sum of contributions of all species,
\begin{equation}
\rho=\rho_{r}+\rho_{b}+\rho_{c}+\rho_{x}\qquad\mbox{and}\qquad p=p_{r}+p_{x}.
\label{rho&p}
\end{equation}
It is convenient to introduce the density parameters
\begin{equation}
\Omega_{i}=\frac{8\pi G}{3H^{2}}\rho_{i},
\label{omega}
\end{equation}
where the index $i$ is running over all components of the universe ($i=r,b,c,x$). Then, Friedmann's equation can be rewritten as
\begin{equation}
\Omega_{r}+\Omega_{b}+\Omega_{c}+\Omega_{x}=1.
\label{friedmann1}
\end{equation}

Radiation and baryons are assumed to evolve independently, their energy densities are given by
\begin{eqnarray}
&\dot{\rho}_{r}+4H\rho_{r}=0\qquad &\Rightarrow\qquad\rho_{r}=\rho_{r0}\ a^{-4},\label{rhor}\\
&\dot{\rho}_{b}+3H\rho_{b}=0\qquad &\Rightarrow\qquad\rho_{b}=\rho_{b0}\ a^{-3}.\label{rhob}
\end{eqnarray}

Since the nature of the dark sector is unknown, we consider a phenomenological interaction  via $T^{\mu\nu}_{c\ ;\nu}=-T^{\mu\nu}_{x\ ;\nu}=Q^{\mu}$, where $Q^{\mu}$ is a four-vector and $T^{\mu\nu}_{c}$ and $T^{\mu\nu}_{x}$ are the energy-momentum tensors of CDM and DE, respectively.
Because of our perfect-fluid description of CDM and DE, the spatial component of the covariant derivative of the energy-momentum tensor must be identically zero, which means that the background interaction term is characterized only by a scalar function $Q$, such that $Q^{\mu}=Qu^{\mu}$. Then, the background energy conservation becomes
\begin{eqnarray}
\dot{\rho}_{c}+3H\rho_{c}=-Q, \label{energycdm}\\
\dot{\rho}_{x}=Q.\label{energyde}
\end{eqnarray}
These equations can be understood as an energy transfer between the dark components. The direction of the energy flux depends on the sign of the scalar function $Q$. For $Q>0$ we have a process of decaying CDM and DE creation, for $Q<0$ the opposite occurs.

Here we are interested in interactions of the type $Q=3H\gamma R\left(\rho_{c},\rho_{x}\right)$, where $\gamma$ is a dimensionless constant and $R$ is a real function with dimension of an energy density. Using this interaction term, the energy balance equations (\ref{energycdm}) and (\ref{energyde}) become
\begin{eqnarray}
\dot{\rho}_{c}+3H\rho_{c}\left(\gamma\dfrac{R}{\rho_{c}}+1\right)=0\,,\label{cdmenergy1}\\
\dot{\rho}_{x}-3\gamma H R=0\,.\label{deenergy1}
\end{eqnarray}
Note that, since $R\left(\rho_{c},\rho_{x}\right)$ is a general function of $\rho_{c}$ and $\rho_{x}$, these equations are coupled.
Now it is convenient to introduce the ratio $r$ of the energy densities of CDM and DE and to consider the time evolution of this quantity,
\begin{equation}
r\equiv\dfrac{\rho_{c}}{\rho_{x}}\qquad\Rightarrow\qquad\dot{r}=r\left(\dfrac{\dot{\rho}_{c}}{\rho_{c}}-\dfrac{\dot{\rho}_{x}}{\rho_{x}}\right).
\label{ratio}
\end{equation}
Combining the equations (\ref{cdmenergy1}), (\ref{deenergy1}) and (\ref{ratio}) one obtains a differential equation for $r$,
\begin{equation}
\dot{r}+3Hr\left(\gamma R\ \dfrac{\rho_{c}+\rho_{x}}{\rho_{c}\, \rho_{x}}+1\right)=0.
\label{dr}
\end{equation}

Equation (\ref{dr}) can be used to decouple equations (\ref{cdmenergy1}) and (\ref{deenergy1}) in case there  exists an analytical solution $r=r\left(a\right)$. Under this condition one may find analytical solutions for $\rho_{c}$ and $\rho_{x}$. To this purpose we require that the first term in the parenthesis of
(\ref{dr}) is a function only of the ratio $r$, i.e.,
\begin{equation}
f\left(r\right)\equiv R\,\dfrac{\rho_{c}+\rho_{x}}{\rho_{c}\,\rho_{x}}\,.
\label{f}
\end{equation}
Using the structure (\ref{f}), equation (\ref{dr}) can be rewritten as
\begin{equation}
\dot{r}+3Hr\left[\dfrac{}{}\gamma\,f\left(r\right)+1\dfrac{}{}\right]=0\,.
\label{rdot}
\end{equation}
Note that the solution of equation (\ref{rdot}) is directly related to the cosmic coincidence problem (CCP) \cite{Velten:2014nra}. Any non-vanishing interaction will modify the ratio $r$ compared to its dependence $r\propto a^{-3}$ within the $\Lambda$CDM model which is recovered for $\gamma=0$. Interacting models have frequently been used to address the CCP (see, e.g., \cite{Chimento:2003iea,Zimdahl:2004hk,Nozari:2014zra}). As we shall show below, the behavior of the solution for $r$ at $a\ll 1$ can also be used to put constraints on the interaction strength.

Here, we are interested in the class of models for which equation (\ref{rdot}) has an analytical solution $r=r\left(a\right)$. In this case the interaction term $R\left(\rho_{c},\rho_{x}\right)$ can be written in terms of only one of the energy densities and the scale factor,
\begin{equation}
R=\dfrac{f\left(r\right)}{1+r}\rho_{c}\qquad\mbox{or}\qquad R=\dfrac{f\left(r\right)}{1+r^{-1}}\rho_{x}.
\label{R}
\end{equation}
Consequently, the energy balance equations (\ref{cdmenergy1}) and (\ref{deenergy1}) of the dark sector become separable,
\begin{eqnarray}
\dot{\rho}_{c}+3H\rho_{c}\left(\gamma\,\dfrac{f\left(r\right)}{1+r}+1\right)=0\,,\label{cdmenergy2}\\
\dot{\rho}_{x}-3\gamma H\rho_{x}\left(\dfrac{f\left(r\right)}{1+r^{-1}}\right)=0\,.\label{deenergy2}
\end{eqnarray}
This encodes the first result of the paper: for interactions resulting in a ratio of the energy densities of CDM and DE which depends only on the scale factor, the individual energy balance equations are always separable.

\textit{A priori}, the function $f\left(r\right)$ can be completely general, but, we assume as an \textit{ansatz} that the interaction term has the following form,
\begin{equation}
Q=3H\gamma\rho_{c}^{\alpha}\rho_{x}^{\beta}\left(\rho_{c}+\rho_{x}\right)^{\sigma}\,,
\label{Q}
\end{equation}
where, on dimensional grounds, the relation $\alpha+\beta+\sigma=1$ must be satisfied. It is straightforward to see that the expression (\ref{Q}) corresponds to $f\left(r\right)=r^{\alpha -1}\left(r+1\right)^{\sigma +1}$. If $\sigma$ is an integer, equation (\ref{Q}) can also be written as a power law using Newton's binomial series,
\begin{eqnarray}
f\left(r\right)=r^{\alpha-1}\qquad &\mbox{ if } \qquad\sigma=-1\,,\label{f1}\\
f\left(r\right)=\sum_{i=0}^{|\sigma+1|}\binom{|\sigma+1|}{i}r^{\alpha-1+i}\qquad &\mbox{ if } \qquad\sigma\neq -1\,.\label{fgeneral}
\end{eqnarray}

The same arguments can be used for interacting DE models with $ w_{x}\neq -1$ for which the inclusion of a factor $\left(w_{x}+1\right)$ in the interaction term has been proposed as a way to avoid instabilities due the DE pressure perturbations \cite{Yang:2017zjs,Yang:2017ccc,Yang:2018euj}. A similar mathematical formulation of interacting models using a function of the ratio between energy densities of CDM and DE can be found in \cite{Chimento:2007yt}.

\subsection{Unified description of interacting models}

An interesting feature of the class of interacting models characterized by (\ref{deenergy1}) is their equivalence  to unified models of the dark sector. In other words, it is possible to combine CDM and DE into a single conservative dark fluid with an EoS
\begin{equation}
p_{d}=w_{d}\left(a\right)\,\rho_{d}\,,
\label{EoS}
\end{equation}
where the subindex $d$ denotes the unified dark fluid.
We define the energy density and the pressure of this dark fluid as a sum of the energy densities and pressures of the components,
\begin{equation}
\rho_{d}=\rho_{c}+\rho_{x}\qquad\mbox{ and }\qquad p_{d}=p_{x},
\end{equation}
respectively. 
The dark-fluid energy density may be written in terms of $r$ and only one of the energy densities $\rho_{c}$
or $\rho_{x}$,
\begin{equation}
\rho_{d}=\left(\dfrac{1+r}{r}\right)\rho_{c}\qquad\mbox{ or }\qquad\rho_{d}=\left(1+r\right)\rho_{x},
\end{equation}
respectively.
Using the second of these options, we conclude that
\begin{equation}
p_{d}=-\dfrac{1}{1+r}\,\rho_{d}\,.
\label{rhod}
\end{equation}
This relation is completely general but the validity of equation (\ref{EoS}) is restricted to interactions for which  the ratio $r$ depends only on the scale factor. In such a case the dark fluid satisfies the conservation equation
\begin{equation}
\dot{\rho}_{d}+3H\rho_{d}\left[1-\dfrac{1}{1+r\left(a\right)}\right]=0.
\label{dark}
\end{equation}
Then, the Hubble rate is obtained through Friedmann's equation using only the unified dark fluid instead of the CDM and DE components separately,
\begin{equation}
H^{2}=\dfrac{8\pi G}{3}\left(\rho_{r}+\rho_{b}+\rho_{d}\right)\,,
\end{equation}
where $\rho_{r}$ and $\rho_{b}$ are, respectively, given by equations (\ref{rhor}) and (\ref{rhob}), and $\rho_{d}$ is the solution of equation (\ref{dark}).
To summarize: in order to describe the background dynamics, there are two equivalent options: the first option is to choose a function $f\left(r\right)$, which means to choose a specific interaction. The second one is to start with  an expression for the ratio $r\left(a\right)$, and to apply equation (\ref{dark}). Since $f\left(r\right)$ and $r\left(a\right)$ are related via equation (\ref{rdot}), specifying only one of these quantities  is sufficient.

\section{Perturbations}
\label{perturbations}

\subsection{Conservation equations}

Restricting ourselves to scalar perturbations in a spatially flat Universe, the perturbed Robertson-Walker metric in the Newtonian gauge with the scalar degrees of freedom $\psi$ and $\phi$ is given by \cite{Ma:1995ey},
\begin{equation}
ds^{2}=a^{2}\left(\tau\right)\left[\dfrac{}{}-\left(1+2\psi\right)d\tau^{2}+\left(1-2\phi\right)dx^{i}dx_{i} \dfrac{}{}\right],
\end{equation}
where, for convenience, the cosmic time $t$ was replaced by the conformal time $\tau$.

In order to describe  structure formation we have to solve the complete set of linear perturbation equations for all components of the Universe. The standard procedure to obtain the CMB temperature anisotropies is to compute the Boltzmann equations for all these components. Here we assume that baryons and radiation behave in the same way as they do in the $\Lambda$CDM model, i.e., interacting with each other via Thomson scattering before recombination but not directly with the dark sector, thus, the Boltzmann equations for these two components will be the same as the well-established equations \cite{Ma:1995ey}. However, since we do not have yet a microscopic description of the interaction between the dark components, corresponding Boltzmann equations are not available either. Instead, we have to use the fluid dynamical description for the components of the dark sector.
Quite generally, the interaction term can be split into components parallel and orthogonal to the four-velocity,
\begin{equation}
Q^{\mu}=Qu^{\mu}+F^{\mu}, \qquad F^{\mu}u_{\mu}=0 \,.
\end{equation}
The background contribution of the scalar function $Q$ already appeared in equations (\ref{energycdm}) and (\ref{energyde}). Writing $Q$ in the covariant form $Q = \Theta\gamma R$, its first-order part, denoted by a hat symbol, is $\hat{Q}=\hat{\Theta}\gamma R+3H\gamma\hat{R}$.  The term $\hat{R}$ depends on the interaction model, i.e., on the energy densities of the dark sector components, the $\hat{\Theta}$ term can be obtained by linearization of the expansion scalar $\Theta\equiv u^{\mu}_{\,;\mu}$ about the homogeneous and isotropic background. In the Newtonian gauge it results in
\begin{equation}
\hat{\Theta}=\dfrac{1}{a}\left(\dfrac{}{}\theta_{tot}-\Theta\psi-3\phi^{\prime}\dfrac{}{}\right).
\end{equation}
Here, $\theta_{tot}\equiv i\,k^{a}\partial_{a}v_{tot}$ where $v_{tot}$ is related to the spatial part of the total four-velocity of the cosmic medium by
$\hat{u}^{\mu}_{tot}=a^{-1}\left(-\psi,\partial^{i}v_{tot}\right)$. The prime denotes a derivative with respect to the conformal time.

The ideal fluid description of the dark components implies that the first-order contribution of $F^{\mu}$ is purely spatial. For the total first-order interaction term we have $\hat{Q}^{\mu}=a\left(Q\psi+\hat{Q},Q\hat{u}^{i}+F^{i}\right)$.

To obtain our basic set of equations we start by considering a general interacting perfect fluid with energy-momentum balance $T^{\mu\nu}_{\ ;\nu} =Q^{\mu}$ and constant EoS $p=w\rho$.
The first-order four-velocity of this fluid is $\hat{u}^{\mu}=a^{-1}\left(-\psi,\partial^{i}v\right)$, where $v$ is its peculiar velocity.
Introducing the density contrast $\delta\equiv\hat{\rho}/\rho$ for this fluid, where $\hat{\rho}$ is its perturbed energy density and $\rho$ is the corresponding background quantity, as well as $\theta\equiv i\,k^{a}\partial_{a}v$,
the well-known energy and momentum conservations in the Newtonian gauge are given by \cite{CalderaCabral:2009ja},
\begin{eqnarray}
\delta^{\prime}+3\mathcal{H}\left(c_{s}^{2}-w\right)\delta+9\mathcal{H}^{2}\left(1+w\right)\left(c_{s}^{2}-c_{a}^{2}\right)\dfrac{\theta}{k^{2}}+\left(1+w\right)\left(\theta-3\phi^{\prime}\right)\nonumber \\
=\dfrac{Qa}{\rho}\left[\dfrac{\hat{Q}}{Q}-\delta+\psi+3\mathcal{H}\left(c_{s}^{2}-c_{a}^{2}\right)\dfrac{\theta}{k^{2}}\right]\,, \label{genenergypert}\\
\theta^{\prime}+\mathcal{H}\left(1-3c_{s}^{2}\right)\theta-\dfrac{k^{2}c_{s}^{2}}{1+w}\,\delta-k^{2}\psi=\dfrac{a}{\rho\left(1+w\right)}\left[\dfrac{}{}Q\theta_{tot}-k^{2}\mathcal{F}\right. \nonumber \\
\left. -\left(1+c_{s}^{2}\right)Q\theta\dfrac{}{}\right]\,.\label{genmomentumpert}
\end{eqnarray}
Here, $\mathcal{H} \equiv \frac{a^{\prime}}{a}$ is the Hubble parameter computed with respect to the conformal time and
$\mathcal{F}$ is defined by $aF^{i}=\partial^{i}\mathcal{F}$.
The quantities $c_{a}^{2}$ and $c_{s}^{2}$ correspond to the squared adiabatic  and  physical rest-frame sound speeds, respectively, of the fluid. Since we consider a constant EoS parameter, the adiabatic sound speed square $c_{a}^{2}$ coincides with $w$. In order to avoid instabilities, the physical sound speed square $c_{s}^{2}$  of dynamical DE has to  be non-negative, Here, we follow the quintessence motivation \cite{Yang:2017ccc,Majerotto:2009np,CalderaCabral:2009ja} and we assume  $c_{s}^{2}=1$.

Now we apply the general equations (\ref{genenergypert}) and (\ref{genmomentumpert}) to each of the dark components. For the CDM component we have,
\begin{eqnarray}
\dot{\delta}_{c}+\theta{c}-3\dot{\phi}=\dfrac{aQ}{\rho_{c}}\left(\delta_{c}-\dfrac{\hat{Q}}{Q}-\psi\right)\,,\label{deltac}\\
\dot{\theta}_{c}+\mathcal{H}\theta_{c}-k^{2}\psi=\dfrac{aQ}{\rho_{c}}\left(\theta_{tot} - \theta_{c}\right)\,.\label{thetac}
\end{eqnarray}
Since $w_{x}=-1$, the velocity $\theta_{x}$ of the DE component has no dynamics. The energy balance is
\begin{equation}
\dot{\delta}_{x}+3\mathcal{H}\left(c_{s}^{2}+1\right)\delta_{x}
=-\dfrac{aQ}{\rho_{x}}\left(\delta_{x}-\dfrac{\hat{Q}}{Q}\right).\label{deltade}
\end{equation}
Note that even if $w_{x}=-1$ the DE component agglomerates. According the equation (\ref{deltade}) fluctuations  of the DE component can have two sources:
the first one is the non-adiabatic character of interacting DE, which leads to a physical sound speed different from the adiabatic sound speed (in this
case, different from -1). Indeed, the non-adiabaticity can play an important role at the linear level \cite{Velten:2018bws}. The second one is the
interaction term on the right-hand side of  equation (\ref{deltade}). In order to solve the set of equations
(\ref{deltac}), (\ref{thetac}) and (\ref{deltade}), we use the well-established adiabatic initial conditions for
interacting models \cite{Majerotto:2009np}.

\section{Specific interacting DE models}
\label{specific}

In the most general case, equation (\ref{dr}) with (\ref{f1}) and (\ref{fgeneral}) has no tractable analytical solution.
For specific  functions $f\left(r\right)$, however, solutions can be found. For some simple choices of  $f\left(r\right)$ we shall recover models that have been previously studied in the literature. The interacting models will be called as IDEM (Interacting Dark Energy Model) followed by a number that will identify each model.

\subsection{IDEM 1: $f\left(r\right)=1$}

The simplest non-vanishing function is $f\left(r\right)=1$, which corresponds to the case $\alpha=\beta=1$ in (\ref{f1}). This parametrization leads to an interaction term
\begin{equation}
Q=3H\gamma \dfrac{\rho_{c}\,\rho_{x}}{\rho_{c}+\rho_{x}}.
\label{q1}
\end{equation}
This interaction term coincides exactly with that of a decomposed generalized Chaplygin gas model \cite{Wang:2013qy,Amendola:2003bz,Marttens:2017njo}.
With (\ref{q1}) equation (\ref{rdot}) for $r$ can be solved to yield
\begin{equation}
r\left(a\right)=r_{0}\ a^{-3\left(\gamma+1\right)},
\label{r1}
\end{equation}
which recovers the corresponding expression in \cite{Funo:2014poa}. Note that, with the reasonable physical assumption that the interaction it is not too strong, i.e., $|\gamma|< 1$, the asymptotic behavior of $r\left(a\right)$ is the same as in the $\Lambda$CDM model: if $a\rightarrow 0$ then $r\left(a\right)\rightarrow \infty$, which means a CDM domination over DE at early times;  if $a\rightarrow \infty$ then $r\left(a\right)\rightarrow 0$, which means a DE domination over CDM in the far future.

\begin{figure}[h!]
\centering
\includegraphics[scale=0.5]{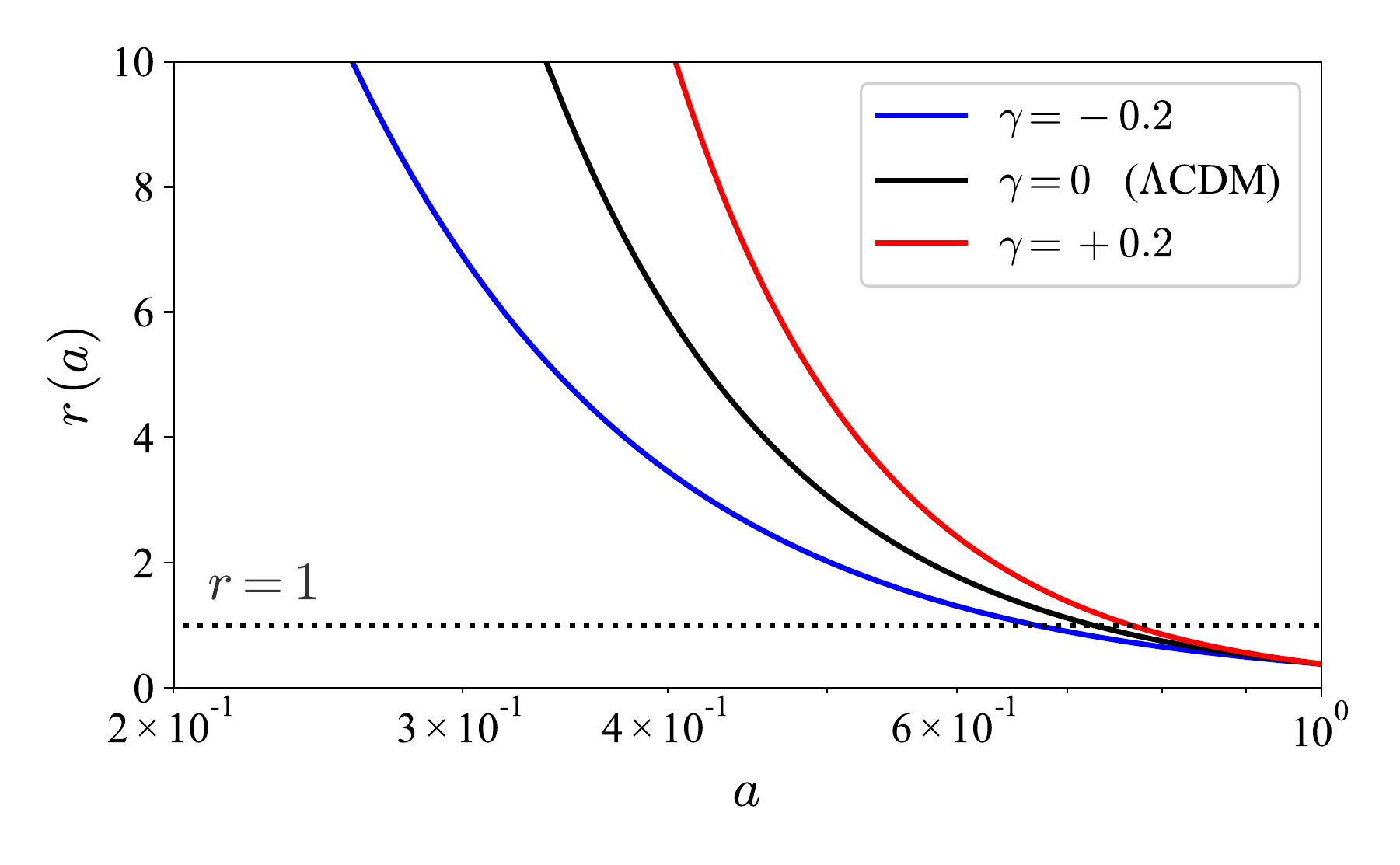}
\caption{Ratio $r\left(a\right)$ between CDM energy density and DE energy density for IDEM 1. }
\label{idem1_r}
\end{figure}

Figure \ref{idem1_r} shows the solution of $r\left(a\right)$ for different values of $\gamma$. For negative values of  the
interaction parameter ($\gamma<0$) the ratio $r\left(a\right)$ reaches the order of $1$ earlier than in the standard model ($
\gamma = 0$). In this sense, the CCP may be considered  alleviated for $\gamma<0$.

The background solutions for the CDM and DE energy densities can be obtained solving the equations (\ref{cdmenergy1}) and (\ref{deenergy1}), which leads to,
\begin{eqnarray}
\rho_{c}=\rho_{c0}\ a^{-3}\left(\dfrac{\Omega_{c0}\ +\Omega_{x0}\ a^{3\left(\gamma+1\right)}}{\Omega_{c0}+\Omega_{x0}}\right)^{-\frac{\gamma}{\gamma+1}},\label{rhoc1}\\
\nonumber \\
\rho_{x}=\rho_{x0}\ a^{-3\left(1+\gamma+1\right)}\left(\dfrac{\Omega_{c0}\ +\Omega_{x0}\ a^{-3\left(\gamma+1\right)}}{\Omega_{c0}+\Omega_{x0}}\right)^{-\frac{\gamma}{\gamma+1}}.\label{rhox1}
\end{eqnarray}
With these energy densities of the dark sector components, Friedmann's equation (\ref{friedmann}) provides us with the Hubble rate square
\begin{equation}
H^{2}=H_{0}^{2}\left[\left(\Omega_{c0}+\Omega_{x0}\right)\left(\dfrac{\Omega_{c0}+\Omega_{x0}\,a^{3\left(\gamma+1\right)}}{\Omega_{c0}+\Omega_{x0}}\right)^{\frac{1}{1+\gamma}}a^{-3}+\Omega_{b0}\,a^{-3}+\Omega_{r0}\,a^{-4}\right]\,.
\label{h1}
\end{equation}

\begin{figure}[h!]
\centering
\includegraphics[scale=0.5]{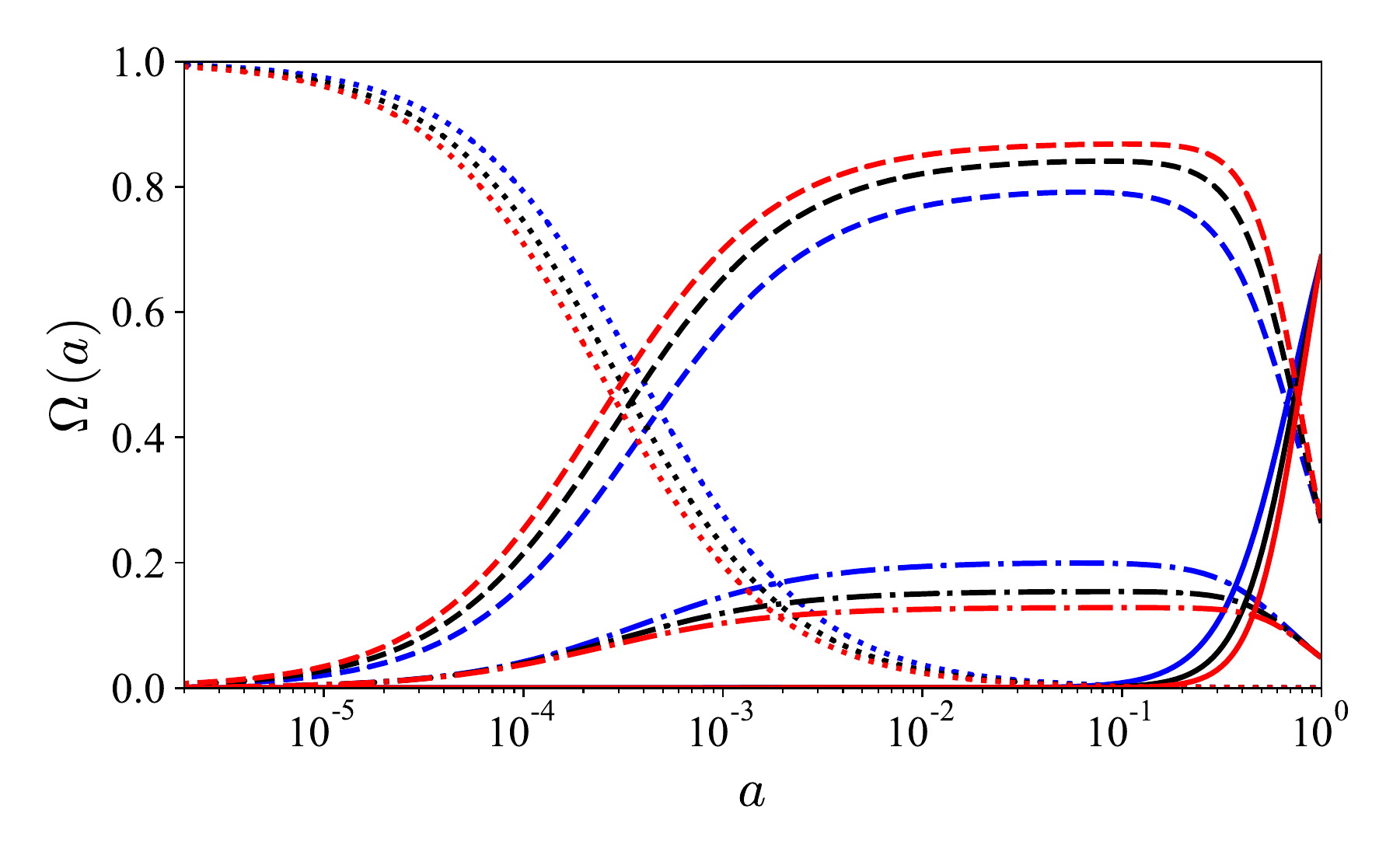}
\caption{Density parameters for all components of the Universe for IDEM 1. The solid lines correspond to the DE component, the dashed lines to CDM, the dot-dashed lines to the baryonic component and the dotted lines to radiation.
Different interaction parameters are distinguished by different colors: blue for to $\gamma=-0.2$, black for $\gamma=0$ (non-interacting case), and red for $\gamma=+0.2$.}
\label{idem1_rho}
\end{figure}

In order to quantify the effect of the interaction on the background solutions it is convenient to analyze the density parameter $\Omega_{i}\left(a\right)=\rho_{i}/\rho_{cr}$, where $\rho_{cr}$ is the critical density, defined as $\rho_{cr}=3H^{2}/8\pi G$. Figure \ref{idem1_rho} shows the density parameters for all components of the Universe using different values of $\gamma$. According to figure \ref{idem1_rho}, negative values for the interaction parameter ($\gamma<0$) delay the equivalence between radiation and matter (CDM + baryons). They also reduce the CDM component and increase the baryonic component during matter domination. Positive values for the interaction parameter ($\gamma>0$) do the opposite. The existence of DE perturbations and the shift of the era of equivalence have
notable and well-known impacts on the physics of the CMB anisotropies.
While the DE perturbations mainly affect large scales of the CMB spectrum, a change
of the era of equivalence considerably alters the radiation driving of the acoustic peaks
and changes the balance between CDM and baryonic matter, leading to a different baryon loading \cite{Amendola:2003bz,Marttens:2017njo}.
In addition, it
is expectable that a non-vanishing $\gamma$ affects the distribution of matter inhomogeneities since
the time of equivalence between radiation and matter is directly related to the location of the peak of the linear
matter power spectrum and to the BAO imprint on it.

\begin{figure}[h!]
\centering
\includegraphics[scale=0.5]{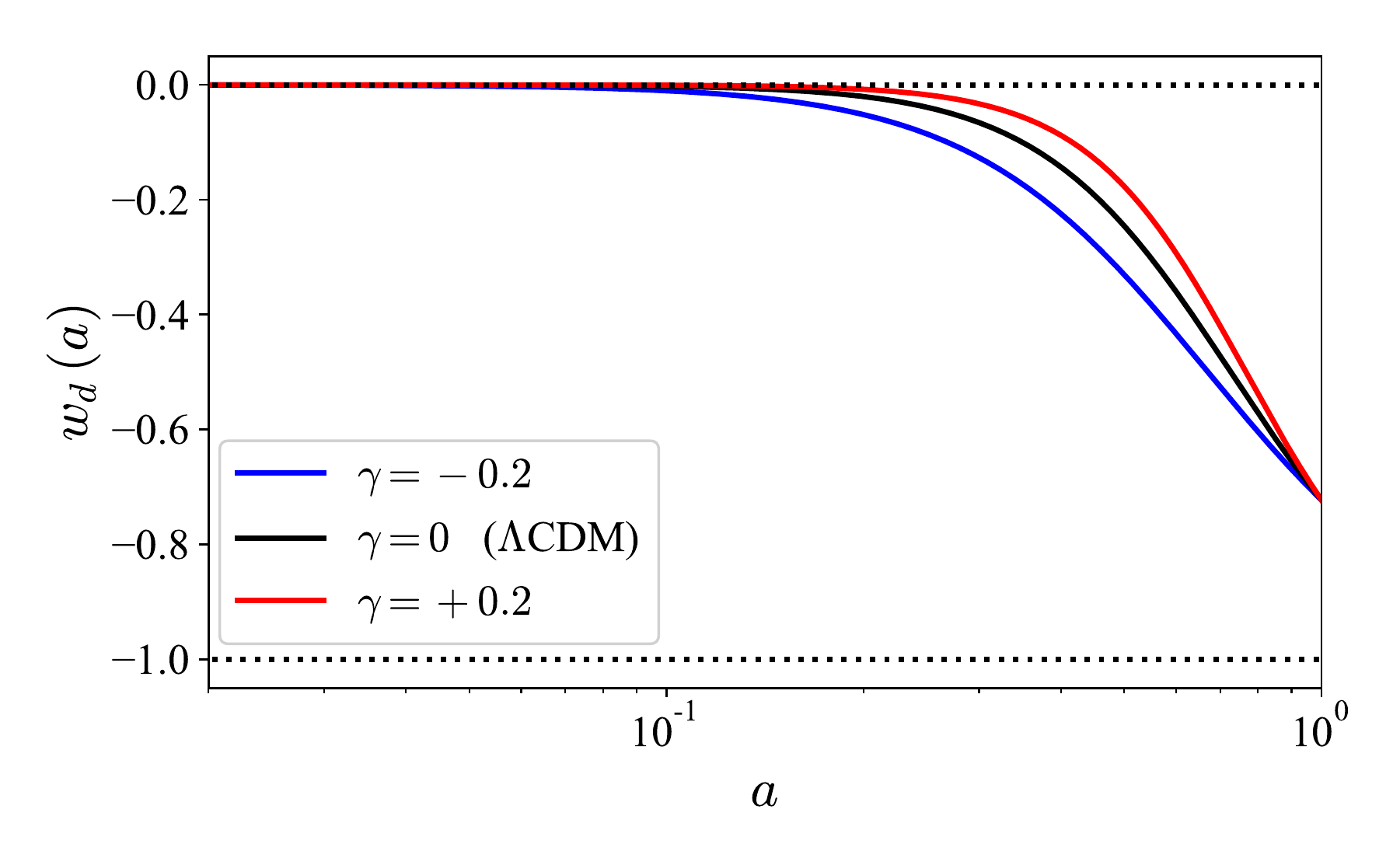}
\caption{EoS parameter for the unified description of  IDEM 1.}
\label{idem1_w}
\end{figure}

The equivalent unified dark-sector model is described through the quantities $w_{d}$ and $\rho_{d}$, for which we find
\begin{equation}
w_{d}\left(a\right)=-\dfrac{\Omega_{x0}}{\Omega_{x0}+\Omega_{c0}\,a^{-3\left(\gamma+1\right)}}
\qquad \mathrm{and} \qquad\rho_{d}=\rho_{d0}\,a^{-3}\left(\dfrac{a^{3\left(\gamma+1\right)}+r_{0}}{1+r_{0}}\right)^{\frac{1}{1+\gamma}},
\label{wd1}
\end{equation}
respectively.
Figure \ref{idem1_w} shows the evolution of the effective dark EoS parameter for different values of $\gamma$. In the past the unified dark fluid behaves like CDM (when $a\rightarrow 0$ we have $w_{d}\rightarrow 0$), and currently
the value for $w_{d}$ is negative. Since $a\rightarrow\infty$ leads to $r\rightarrow 0$, the dark EoS parameter tends to $w_{d} =-1$ in the far-future limit.

Recalling that in the background the expansion scalar $\Theta$ reduces to  $\Theta = 3H$, the interaction term (\ref{q1}) can be seen as a covariant scalar quantity. Then, its first-order perturbation is
\begin{equation}
\hat{Q}=Q\left(\dfrac{\hat{\Theta}}{\Theta}+\dfrac{\rho_{c}\delta_{x}+\rho_{x}\delta_{c}}{\rho_{c}+\rho_{x}}\right)\,.
\label{hq1}
\end{equation}
This completes our description of the IDEM 1.

\subsection{IDEM 2: $f\left(r\right)=\dfrac{1}{r}$}

\begin{figure}[h!]
\centering
\includegraphics[scale=0.5]{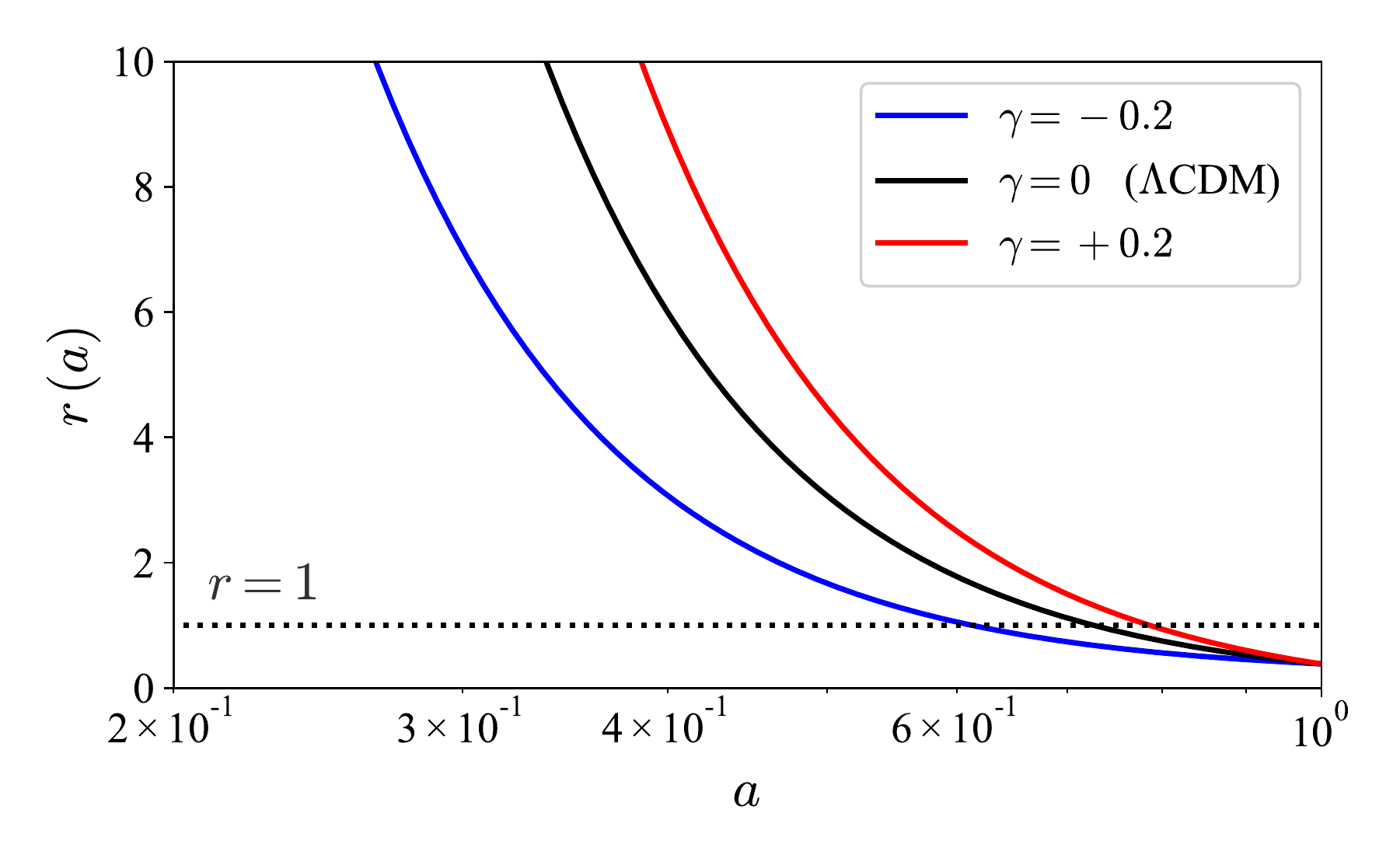}
\caption{Ratio between CDM energy density and DE energy density $r\left(a\right)$ for IDEM 2.}
\label{idem2_r}
\end{figure}

\begin{figure}[h!]
\centering
\includegraphics[scale=0.5]{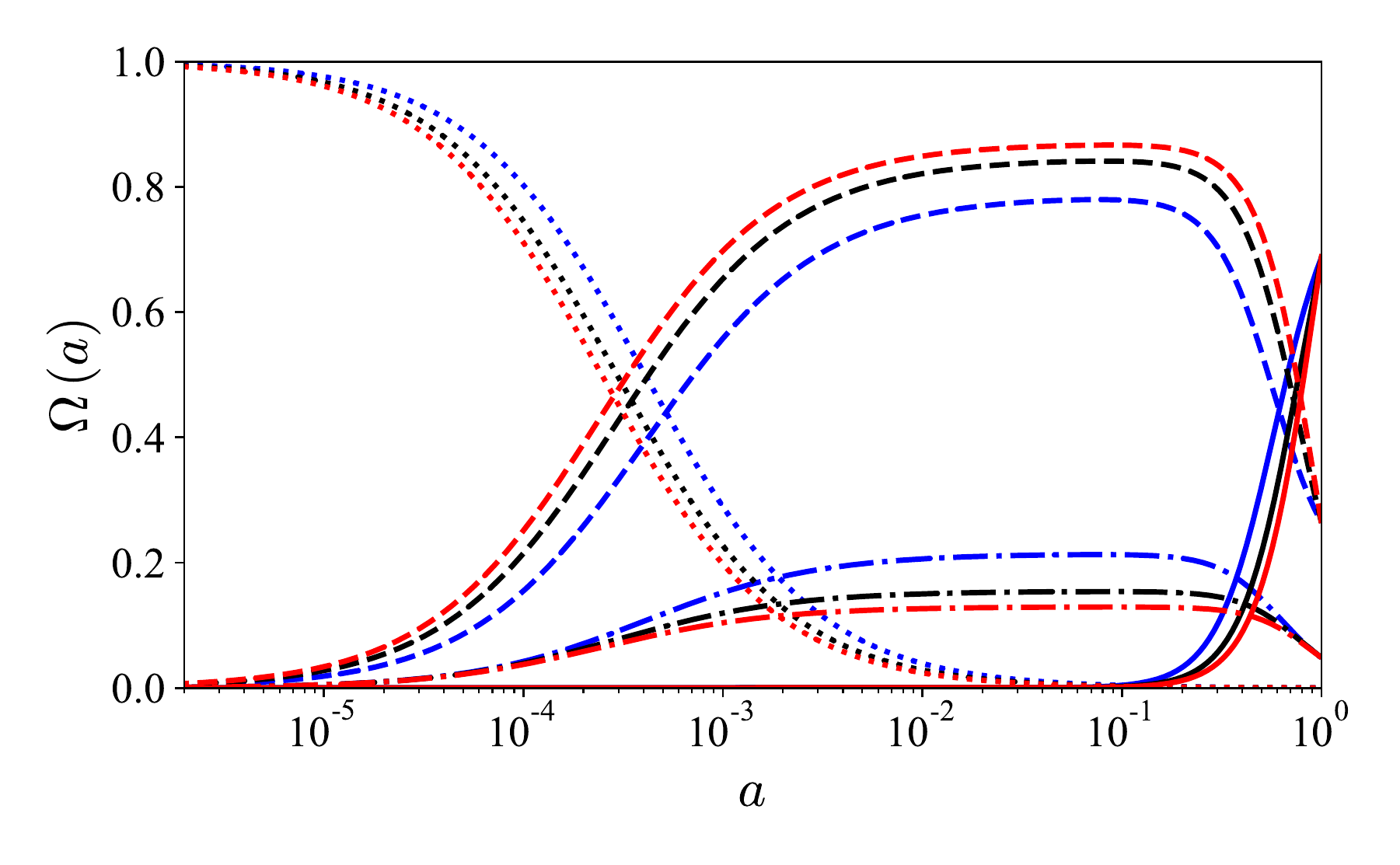}
\caption{Density parameters for all components of the Universe for IDEM 2. The solid lines correspond to DE, the dashed lines to CDM, the dot-dashed lines to baryons and the dotted lines to radiation. Curves in blue refer to $\gamma=-0.2$, curves in black  to $\gamma=0$ (non-interacting case) curves in red to $\gamma=+0.2$.}
\label{idem2_rho}
\end{figure}

The second case studied is $f\left(r\right)=1/r$. It is obtained from (\ref{f1}) with $\alpha=0$ and $\beta=2$, equivalent to an interaction term
\begin{equation}
Q=3H\gamma\dfrac{\rho_{x}^{2}}{\rho_{c}+\rho_{x}}.
\label{q2}
\end{equation}
A statistical analysis of this model using SNe Ia data was performed in \cite{Arevalo:2011hh}.
From equation (\ref{dr}) we obtain for $r\left(a\right)$,
\begin{equation}
r\left(a\right)=r_{0}\,a^{-3}-\gamma\left(1-a^{-3}\right)\,.
\label{r2}
\end{equation}
This solution has an interesting asymptotic behavior. In the early universe, when $a$ tends to zero, the ratio between CDM and DE energy densities goes to $\left(r_{0}+\gamma\right)a^{-3 }$
i.e., it diverges.  This limit means that  CDM always dominates over DE in the past, but, if $\gamma<-r_{0}$, the DE density arises from an initial negative regime. In the following we shall exclude such primordial negative DE density phase by imposing the constraint $\gamma>-r_{0}$.
 In the far future, i.e., when $a\gg 1$, solution \ref{r2} tends to $-\gamma$, i.e., a certain amount of CDM will persist forever. Furthermore, this limit implies that for a positive interaction parameter the DE density will become negative in the future. Figure \ref{idem2_r} shows the ratio $r\left(a\right)$ for IDEM 2 for different values of $\gamma$. Again, negative values of the interaction parameter ($\gamma<0$) can alleviate the CCP.
Equation (\ref{deenergy2}) provides us with the background DE energy density
\begin{equation}
\rho_{x}=\rho_{x0}\ a^{-\frac{3\gamma}{\gamma-1}}\left[\dfrac{\left(1-\gamma\right)\Omega_{x0}+a^{-3}\left(\Omega_{c0}+\gamma\ \Omega_{x0}\right)}{\left(\Omega_{c0}+\Omega_{x0}\right)}\right]^{-\frac{\gamma}{\gamma-1}}.
\label{rhox2}
\end{equation}
The CDM energy density is found by combining (\ref{rhox2}) with (\ref{r2}). Then, the background dynamics is completely known. Figure \ref{idem2_rho} shows the density parameters for all components for different values of $\gamma$. Obviously, the behavior of the background solutions for IDEM 2 is very similar to that of IDEM 1.

\begin{figure}[h!]
\centering
\includegraphics[scale=0.5]{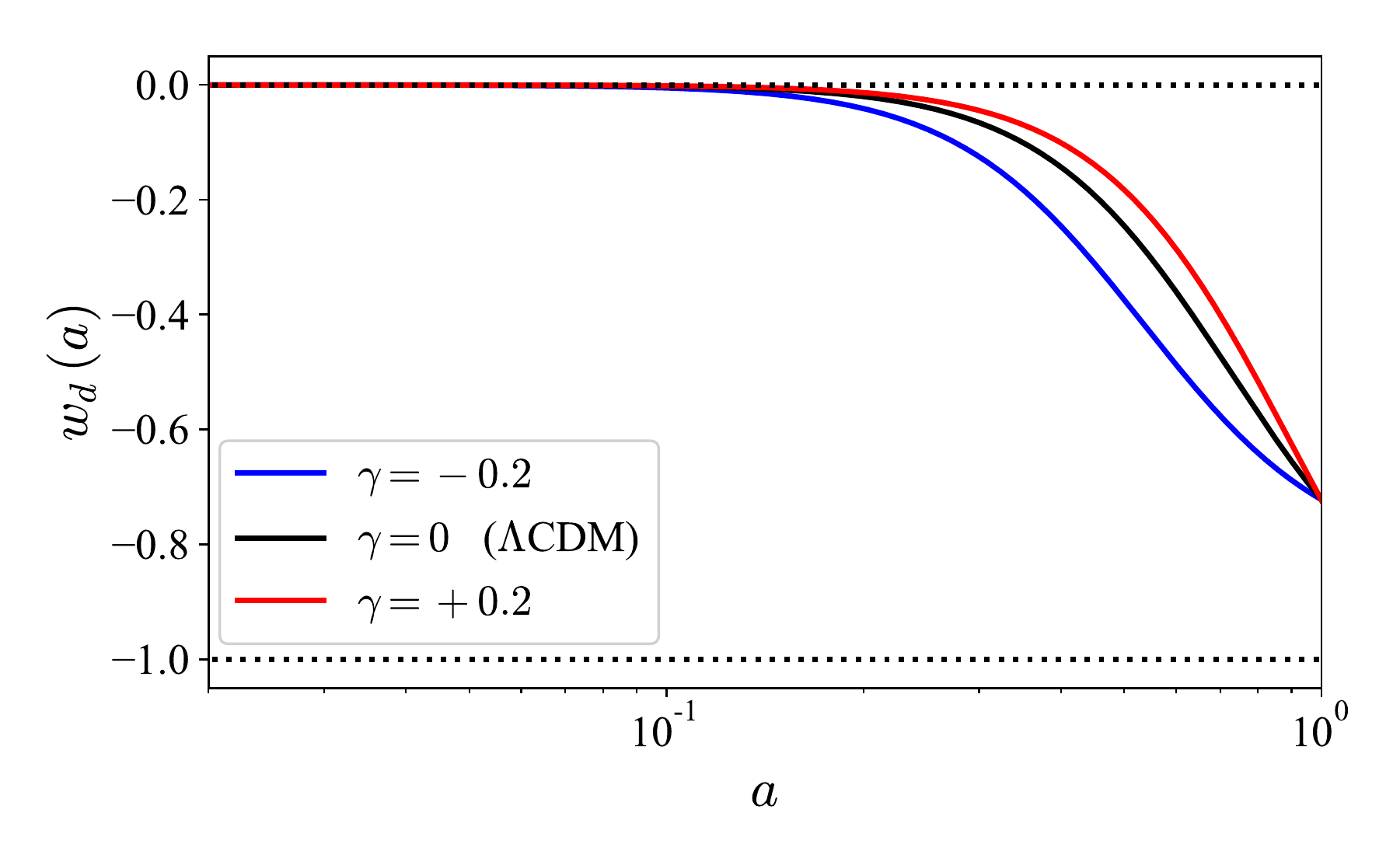}
\caption{EoS parameter for the unified description of model IDEM 2.}
\label{idem2_w}
\end{figure}

The unified model of the dark sector is determined by the combination  of equations (\ref{rhod}) and (\ref{r2}). The effective dark EoS parameter is shown in figure \ref{idem2_w}. It is also very similar to that of model IDEM 1.

Finally, at linear order the interaction parameter  (\ref{q2}) is
\begin{equation}
\hat{Q}=Q\left(\dfrac{\hat{\Theta}}{\Theta}+\dfrac{\rho_{x}\delta_{x}+2\rho_{c}\delta_{x}-\rho_{c}\delta_{c}}{\rho_{c}+\rho_{x}}\right)\,.
\label{hq2}
\end{equation}

\subsection{IDEM 3: $f\left(r\right)=r$}

\begin{figure}[h!]
\centering
\includegraphics[scale=0.5]{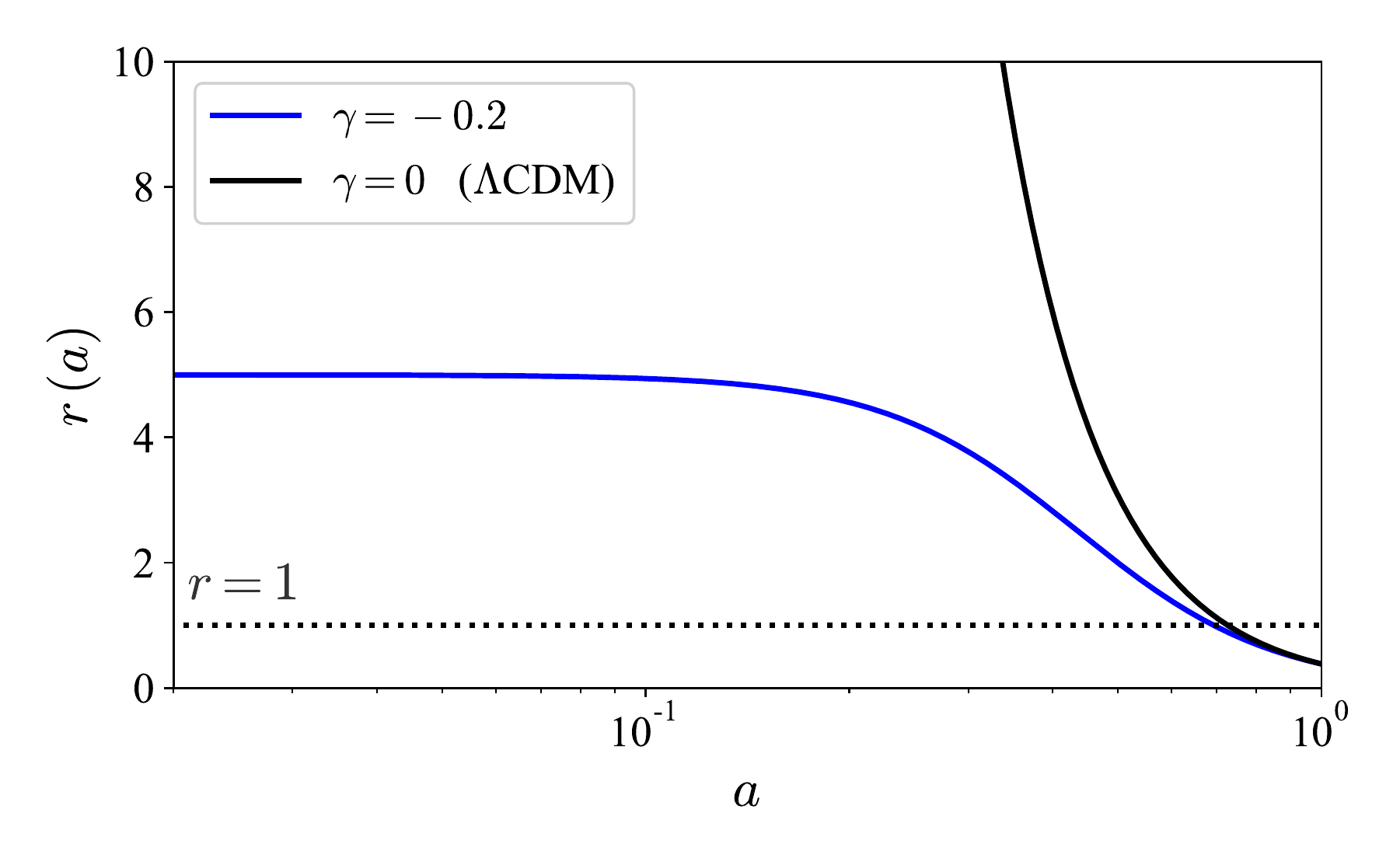}
\caption{Ratio $r\left(a\right)$  between CDM energy density and DE energy for IDEM 3.}
\label{idem3_r}
\end{figure}

\begin{figure}[h!]
\centering
\includegraphics[scale=0.5]{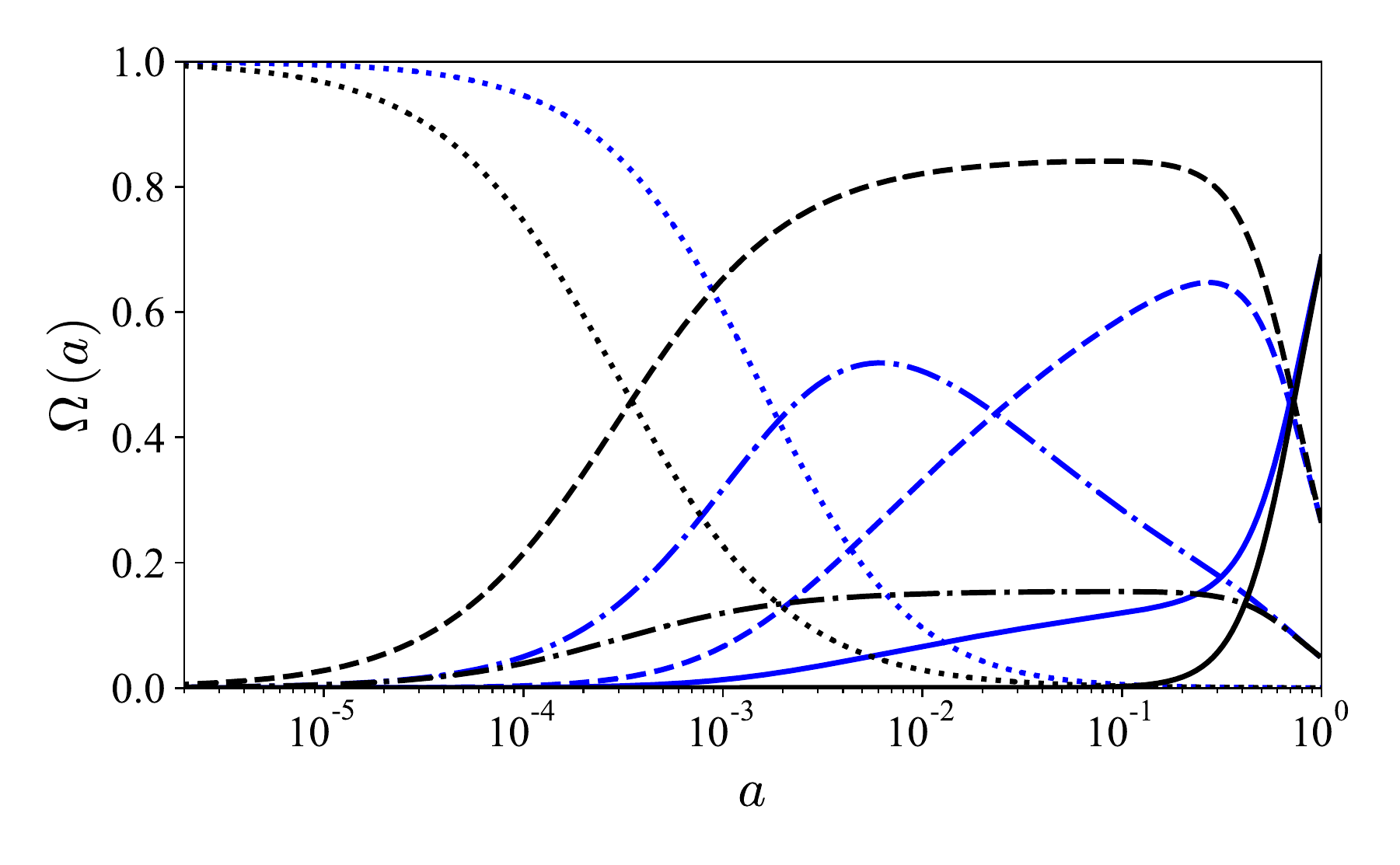}
\caption{Density parameter for all components of the Universe for IDEM 3. The solid lines correspond to DE, the dashed lines
to CDM, the dot-dashed lines to the baryonic component and the dotted lines to radiation. The blue curves  refer to $
\gamma=-0.2$ and the black ones to $\gamma=0$ (non-interacting case).}
\label{idem3_rho}
\end{figure}

The choice $f\left(r\right)=r$ is realized for $\alpha=2$ and $\beta=0$ in (\ref{f1}). It leads to the interaction parameter
\begin{equation}
Q=3H\gamma\dfrac{\rho_{c}^{2}}{\rho_{c}+\rho_{x}}.
\label{q3}
\end{equation}
A statistical analysis with SNe Ia data for this model was also performed in \cite{Arevalo:2011hh}. With $f\left(r\right)=r$ equation (\ref{dr}) yields
\begin{equation}
r\left(a\right)=r_{0}\dfrac{a^{-3}}{1+r_{0}\gamma -r_{0}\gamma a^{-3}}.
\label{r3}
\end{equation}
In the early-universe limit, i.e. for $a\ll 1$, the ration between CDM and DE energy densities tends to $-1/\gamma$. Since we
wish to avoid an early negative DE density phase again, we require the interaction parameter to be negative. For $a\gg 1$,
independently of the value of $\gamma$, the ratio $r\left(a\right)$ tends to zero. Figure \ref{idem3_r} shows $r\left(a
\right)$ for $\gamma = -0.2$ and $\gamma = 0$. It is evident that for a non-vanishing $\gamma < 0$  the ratio between
CDM and DE energy densities does not diverge for $a\ll 1$. In a sense, an interactions of this type may  solve the CCP.
Equation (\ref{cdmenergy2}) for the CDM energy density results in
\begin{equation}
\rho_{c}=\rho_{c0}\ a^{-3}\left[\dfrac{\gamma\Omega_{c0}+\left(1-\gamma\right)\Omega_{c0}a^{-3}+\Omega_{x0}}{\left(\Omega_{c0}+\Omega_{x0}\right)}\right]^{-\frac{\gamma}{\gamma-1}}\,.
\label{rhoc3}
\end{equation}
The corresponding DE energy density follows from  (\ref{rhoc3}) with (\ref{r2}). Figure \ref{idem3_rho} shows the density
parameters for all components for different values of $\gamma$. As one can see, even a small value of the interaction
parameter can modify drastically the background evolution of all components of the universe and, consequently, the
entire expansion history. For this reason it is expectable that the data will strongly constrain the
interaction parameter for IDEM 3.

\begin{figure}[h!]
\centering
\includegraphics[scale=0.5]{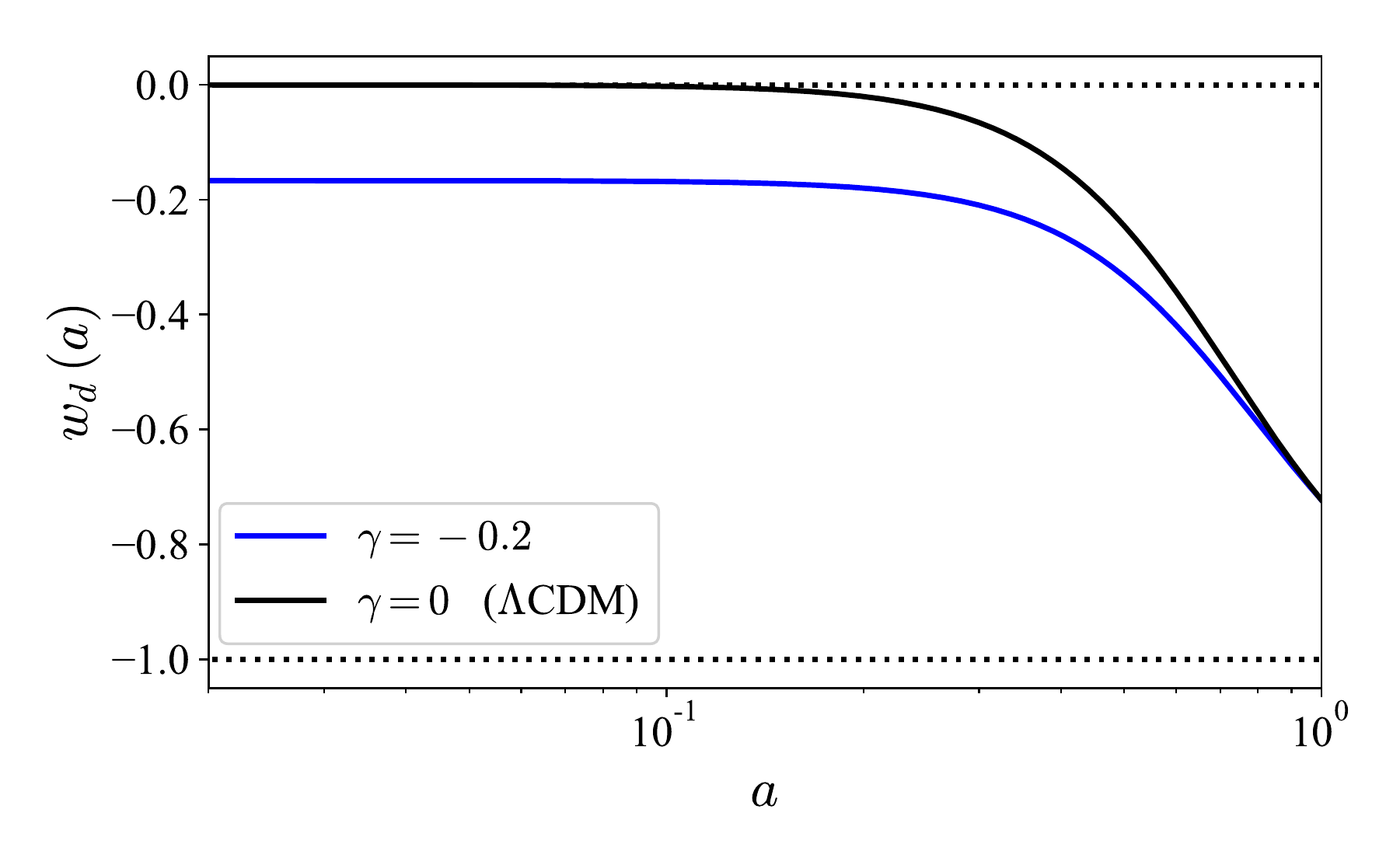}
\caption{EoS parameter for the unified description of model IDEM 3. The blue lines correspond to $\gamma=-0.2$, the black lines correspond to $\gamma=0$ (non-interacting case).}
\label{idem3_w}
\end{figure}

The effective dark EoS parameter in figure \ref{idem3_w} is obtained by combining  equations (\ref{rhod}) and (\ref{r3}). The value of $|\gamma|$ quantifies the deviation from pure matter domination in the early universe.

The first-order  interaction parameter of model IDEM 3 becomes
\begin{equation}
\hat{Q}=Q\left(\dfrac{\hat{\Theta}}{\Theta}+\dfrac{\rho_{c}\delta_{c}+2\rho_{x}\delta_{c}
-\rho_{x}\delta_{x}}{\rho_{c}+\rho_{x}}\right)\,.
\label{hq3}
\end{equation}

\subsection{IDEM 4: $f\left(r\right)=1+\dfrac{1}{r}$}

With $\alpha=0$, $\beta=1$ and $\sigma=0$ in (\ref{fgeneral}) one has $f\left(r\right)=1+1/r$ and
\begin{equation}
Q=3H\gamma\rho_{x}.
\label{q4}
\end{equation}
From (\ref{dr}), one finds the ratio $r\left(a\right)$,
\begin{equation}
r\left(a\right)=a^{-3\gamma}\dfrac{\left(r_{0}a^{-3}+\gamma a^{-3}-\gamma a^{3\gamma}+r_{0}\gamma a^{-3}\right)}{1+\gamma}\,.
\label{r4}
\end{equation}
In the early universe, when $a$ tends to zero, the ratio $r(a)$ diverges, i.e, CDM always dominates.
The sign of $r(a)$ in this limit is determined by the combination $\left[\gamma+r_{0}\left(1+\gamma\right)\right]$.
For $\gamma< -r_{0}/\left(1+r_{0}\right)$ DE arises from a negative DE density regime.
To
exclude an early negative DE density phase we shall restrict ourselves to
$\gamma>-r_{0}/\left(1+r_{0}\right)$ which puts a negative lower bound on
the value of the interaction parameter.
In the far future, i.e., for $a\gg 1$, the solution (\ref{r4}) tends to $-\gamma/\left(1+\gamma\right)$, which means that there is
a remaining CDM component that exists forever. As for IDEM 2, a positive $\gamma$ leads to negative DE density in the future.

\begin{figure}[h!]
\centering
\includegraphics[scale=0.5]{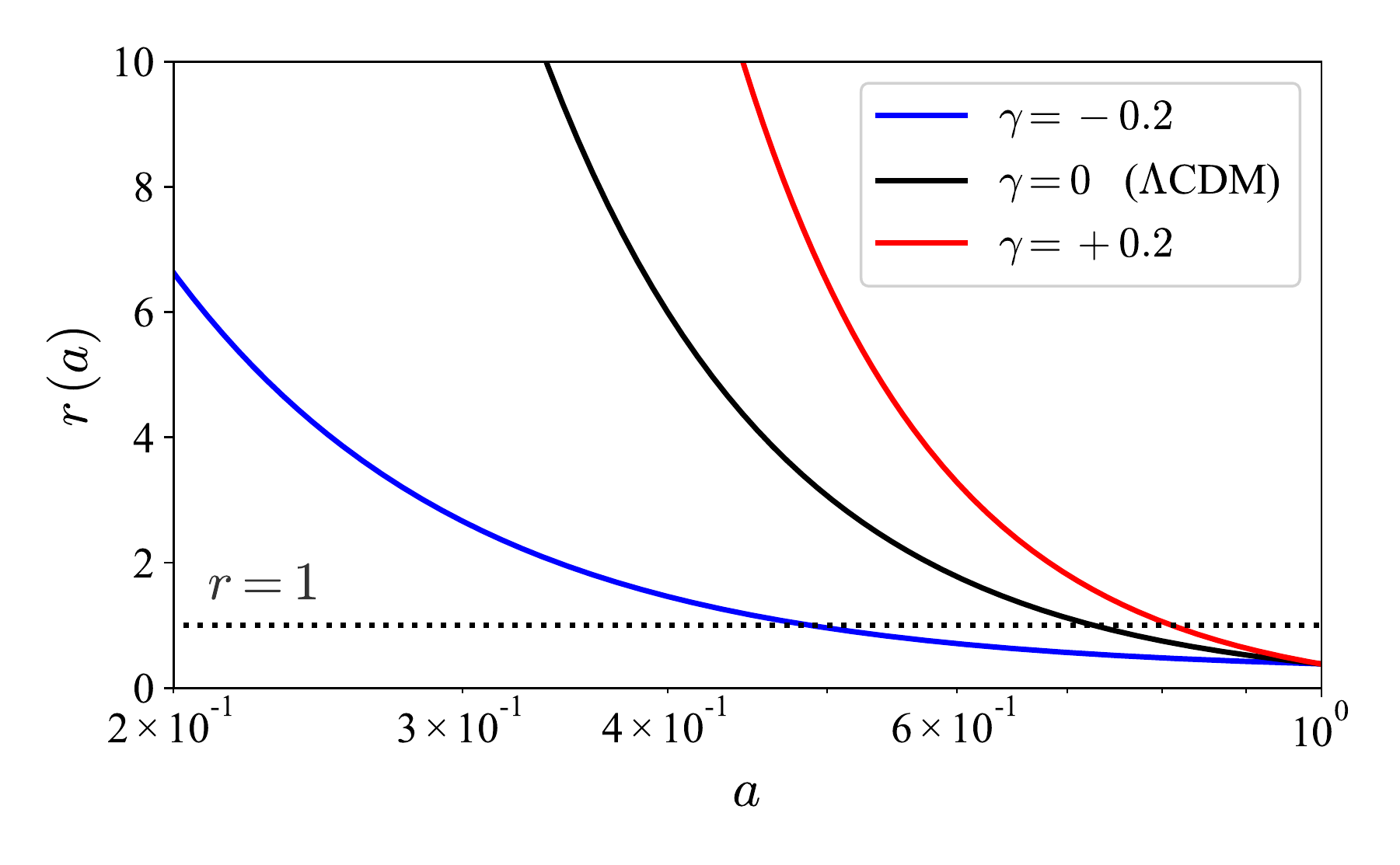}
\caption{Ratio $r\left(a\right)$ between CDM energy density and DE density for IDEM 4.}
\label{idem4_r}
\end{figure}

\begin{figure}[h!]
\centering
\includegraphics[scale=0.5]{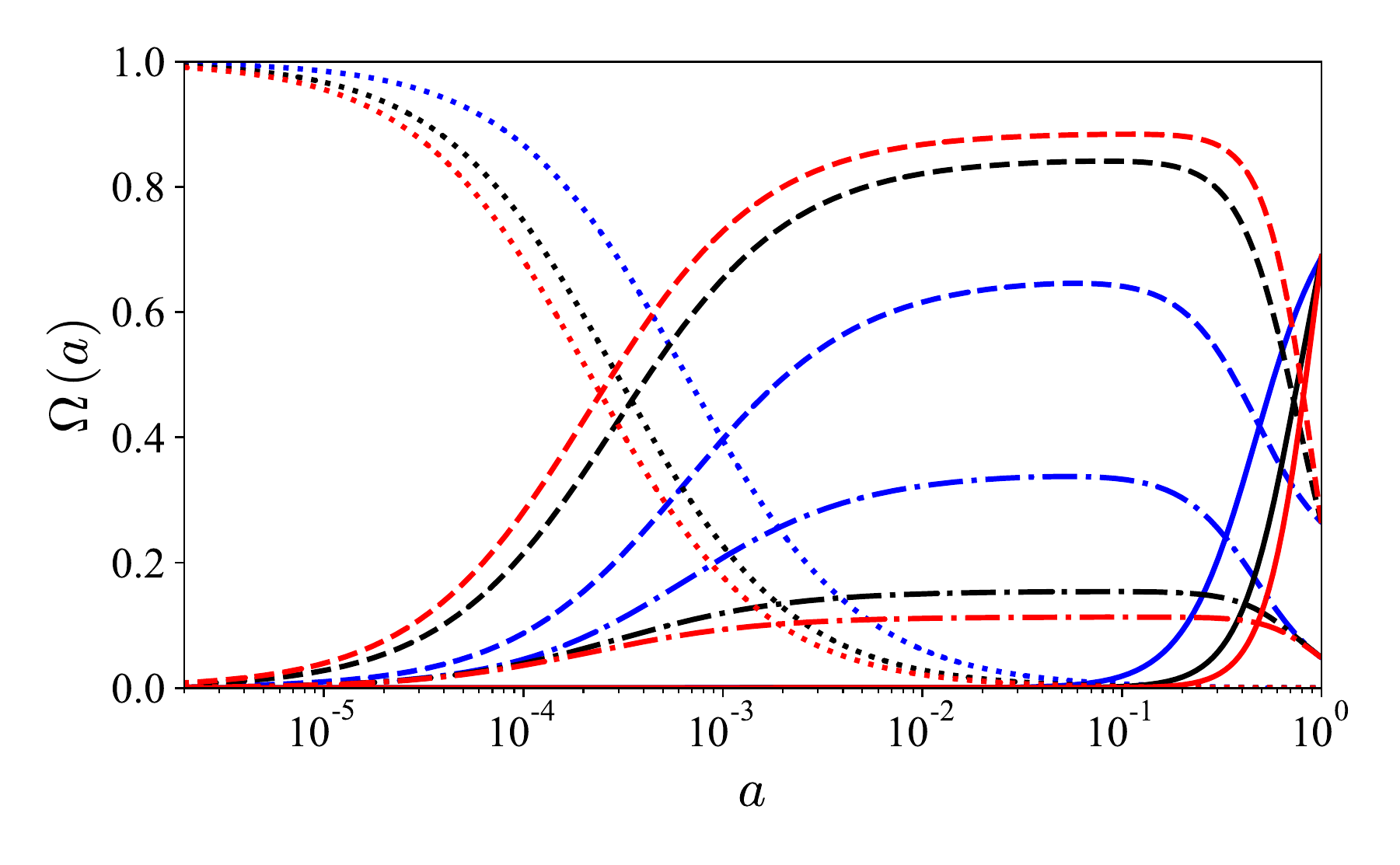}
\caption{Density parameters for all components of the Universe for IDEM 4. The solid lines correspond to DE, the dashed
lines to CDM component, the dot-dashed lines to baryons and the dotted lines to the radiation component. The blue
color denotes the case $\gamma=-0.2$, the black curves refer to $\gamma=0$ (non-interacting case) and the red ones to $
\gamma=+0.2$.}
\label{idem4_rho}
\end{figure}

\begin{figure}[h!]
\centering
\includegraphics[scale=0.5]{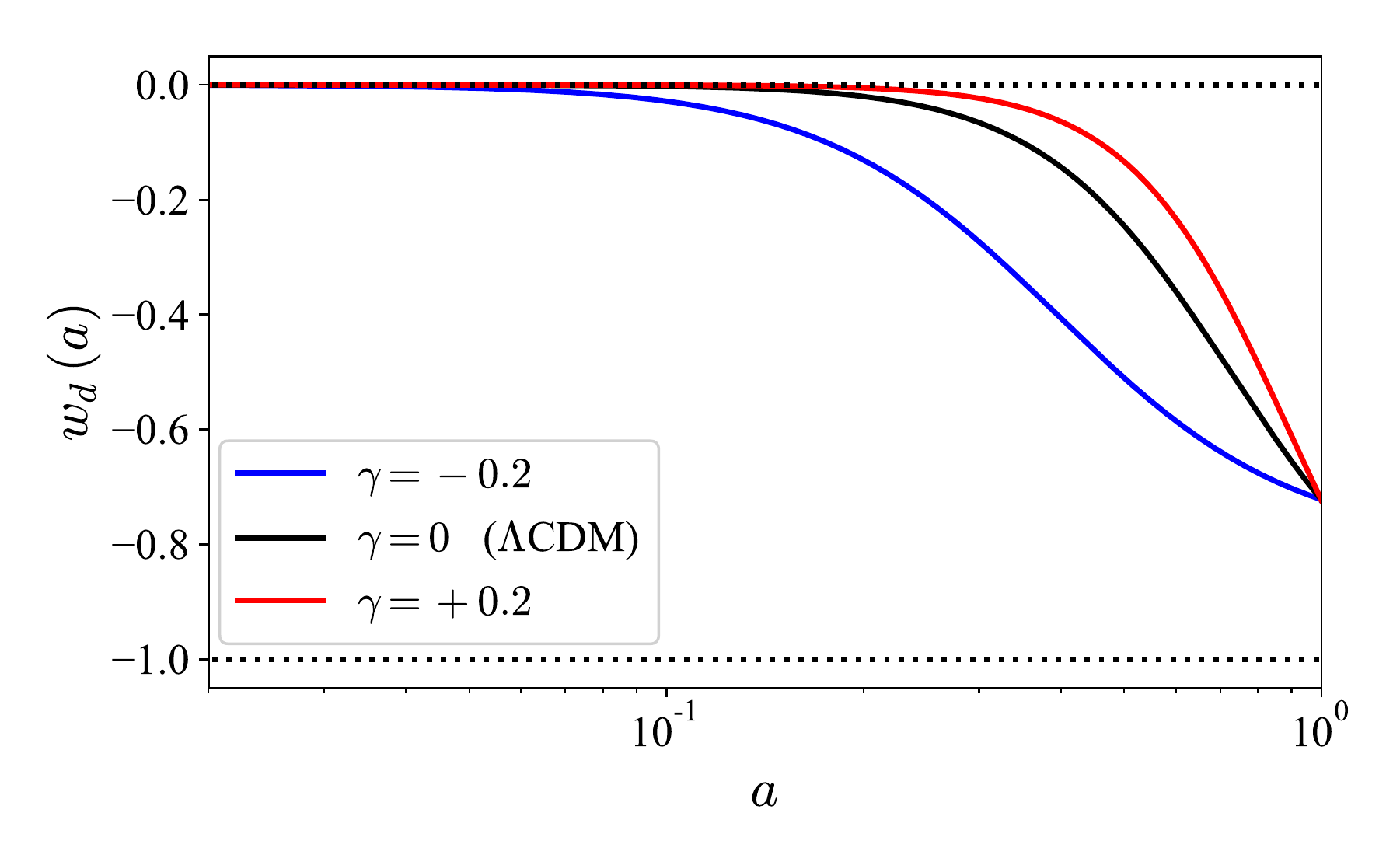}
\caption{EoS parameter for the unified description of IDEM 4. }
\label{idem4_w}
\end{figure}

Figure \ref{idem4_r} shows the ratio $r\left(a\right)$ for the IDEM 4 model for different values of $\gamma$. Just like IDEM 1 and IDEM 2, negative values of the interaction parameter ($\gamma<0$) can alleviate the CCP.
From equation (\ref{deenergy2}) the background DE density for this model is
\begin{equation}
\rho_{x}=\rho_{x0}\ a^{3\gamma}.\label{rhox4}
\end{equation}
Together with (\ref{r4}) it determines the background dynamics. Figure \ref{idem4_rho} shows the density parameters for all
components for different values of $\gamma$. The background dynamics of model IDEM 4 is very similar to the dynamics of
the previously studied models IDEM 1 and IDEM 2.
The effective unified dark fluid EoS parameter which follows from (\ref{rhod}) with (\ref{r4}) is shown in Figure \ref{idem4_w}. It is also very similar to the EoS parameters of models IDEM 1 and IDEM 2.
The  linear-order expression for the interaction parameter, calculated from (\ref{q4}), is
\begin{equation}
\hat{Q}=Q\left(\dfrac{\hat{\Theta}}{\Theta}+\delta_{x}\right)\,.
\label{hq4}
\end{equation}

\subsection{IDEM 5: $f\left(r\right)=1+r$}

\begin{figure}[h!]
\centering
\includegraphics[scale=0.5]{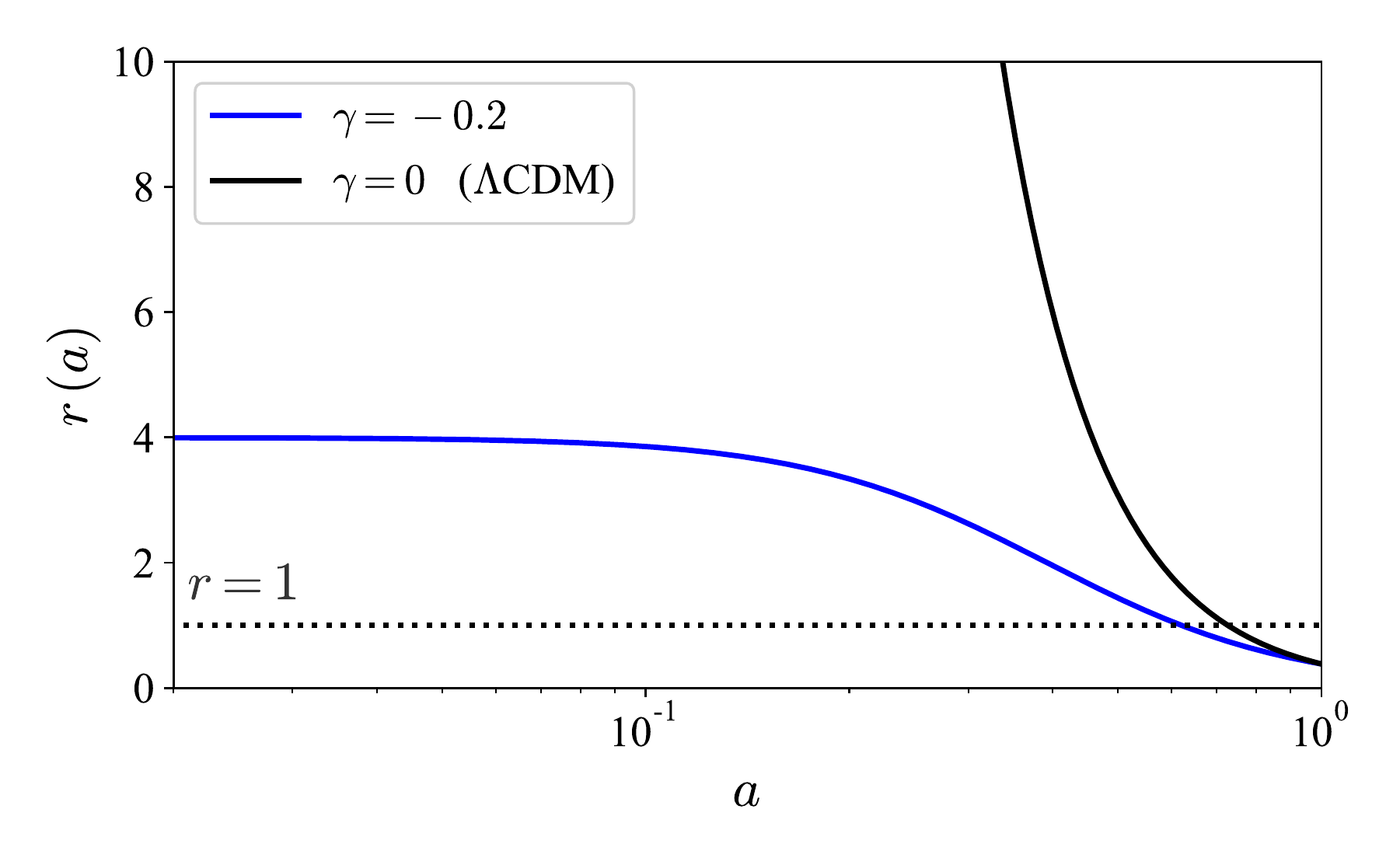}
\caption{Ratio $r\left(a\right)$ between CDM energy density and DE density for IDEM 5. }
\label{idem5_r}
\end{figure}

\begin{figure}[h!]
\centering
\includegraphics[scale=0.5]{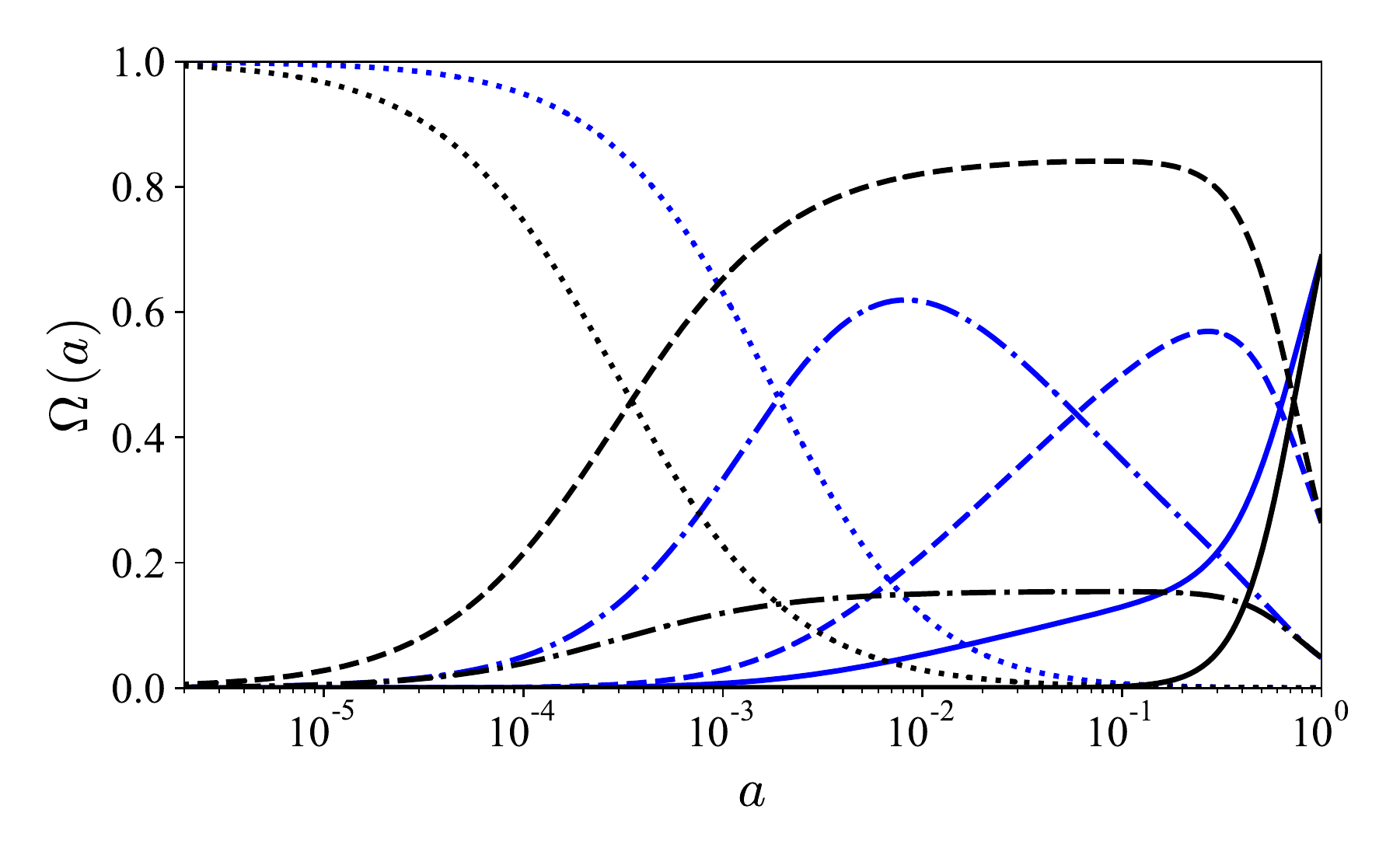}
\caption{Density parameters for all components of the Universe for IDEM 5. The solid lines correspond to DE, the dashed
lines to CDM, the dot-dashed lines to baryonic matter and the dotted lines  to the radiation component. Blue curves  refer to
$\gamma=-0.2$ and  black curves to $\gamma=0$ (non-interacting case).} \label{idem5_rho}
\end{figure}

\begin{figure}[h!]
\centering
\includegraphics[scale=0.5]{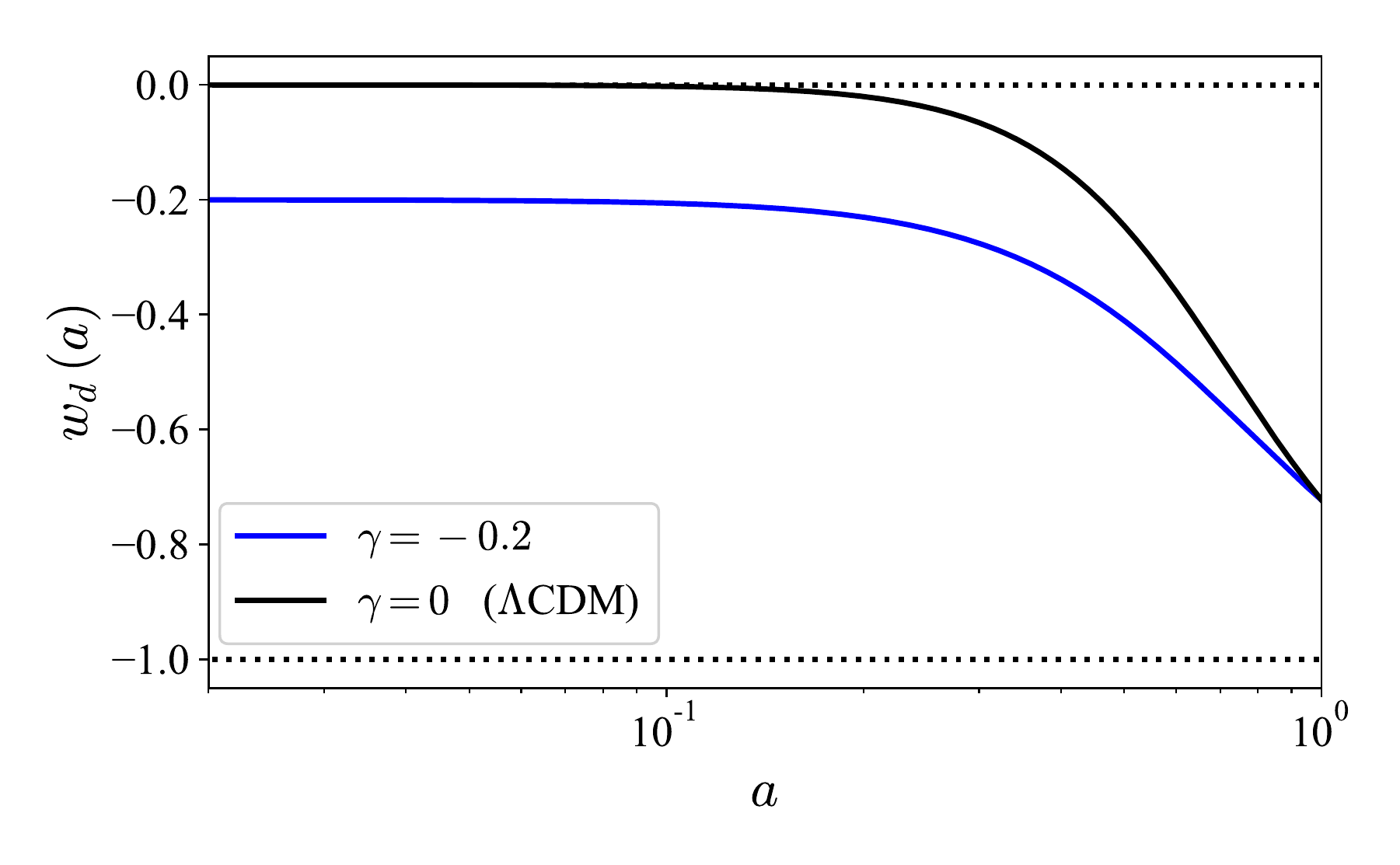}
\caption{EoS parameter for the unified description of IDEM 5. }
\label{idem5_w}
\end{figure}

Equation (\ref{fgeneral}) with $\alpha=1$, $\beta=0$ and $\sigma=0$ results in $f\left(r\right)=1+r$
and in
\begin{equation}
Q=3H\gamma\rho_{c}
\label{q5}
\end{equation}
for the interaction parameter.
The solution of equation \ref{dr} is
\begin{equation}
r\left(a\right)=-\dfrac{1+\gamma}{\gamma-a^{3\left(1+\gamma\right)}\left(\dfrac{1+\gamma+r_{0}\gamma}{r_{0}}\right)}.
\label{r5}
\end{equation}
For $a\ll 1$ the ratio $r$  tends to $-\left(1+\gamma\right)/\gamma$, where the relation $\gamma>-1$ must be satisfied. Excluding again an initial negative DE density phase, we restrict our analysis to negative values of $\gamma$ with $|\gamma|<1$.
For $a\gg 1$  the ratio $r\left(a\right)$ tends to zero. Figure \ref{idem5_r} shows $r\left(a\right)$ for $\gamma=-0.2$ and $\gamma=0$. For $\gamma=-0.2$  the ratio between CDM and DE densities does not diverge for $a\ll 1$ which is a similar feature as already found for model IDEM 3.
Equation (\ref{cdmenergy2}) yields
\begin{equation}
\rho_{c}=\rho_{c0}\ a^{-3\left(1+\gamma\right)},
\label{rhoc5}
\end{equation}
such that the background dynamics is fixed together with (\ref{r2}). Figure \ref{idem5_rho} shows the density parameters for
$\gamma=-0.2$ and $\gamma=0$. As in model IDEM 3, already a small non-vanishing interaction parameter can
substantially modify the background evolution of all the components.
From (\ref{rhod}) and (\ref{r5}) one finds  the effective unified dark EoS parameter visualized in  Figure \ref{idem5_w}. Similar
to model IDEM 3, the absolute value $|\gamma|$ quantifies the difference to the EoS for pressureless matter at $a\ll 1$.
The linearized  interaction parameter for this model is given by
\begin{equation}
\hat{Q}=Q\left(\dfrac{\hat{\Theta}}{\Theta}+\delta_{c}\right)\,.
\label{hq5}
\end{equation}

\subsection{Interaction time evolution}

\begin{figure}[h!]
\centering
\includegraphics[scale=0.37]{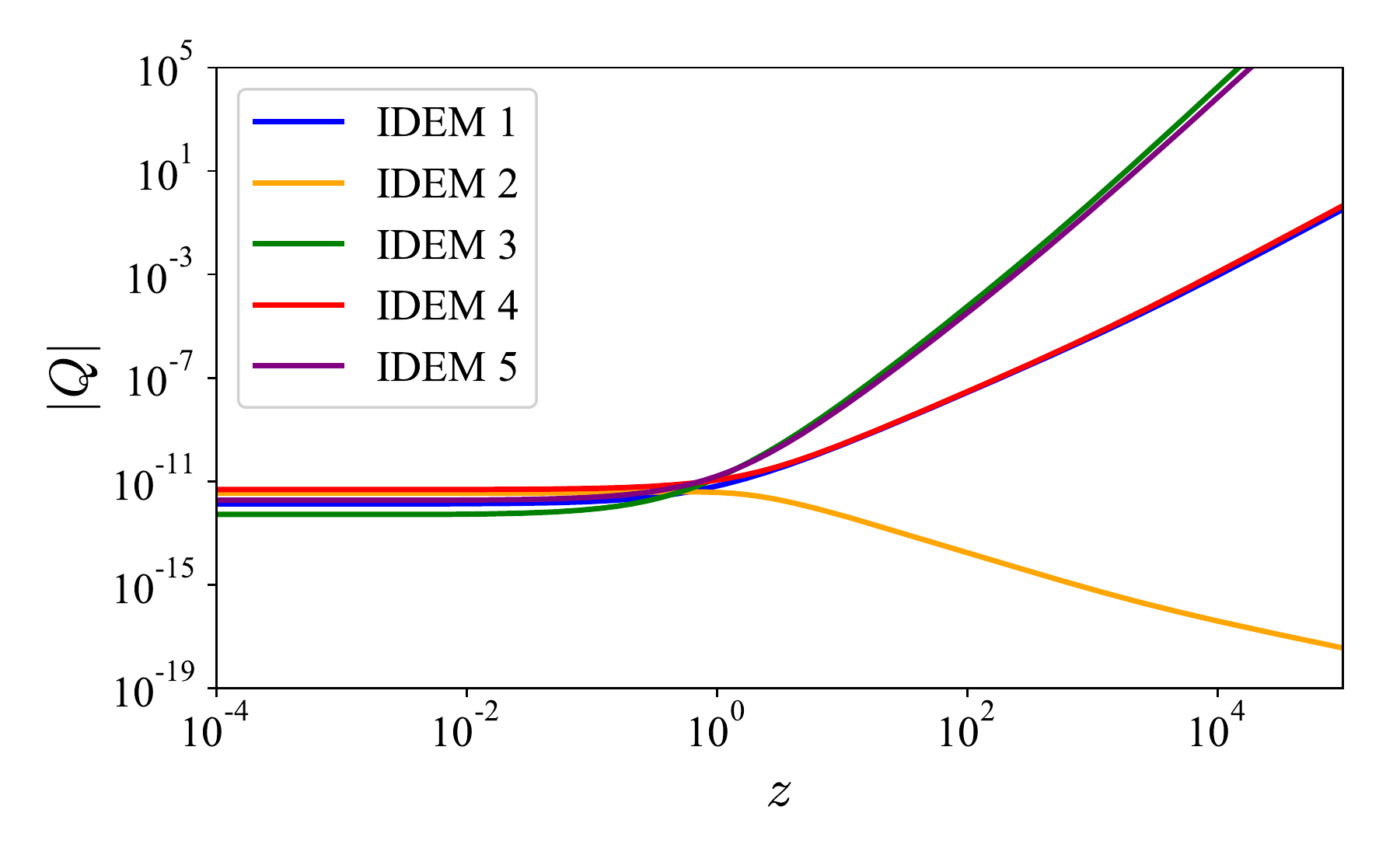}
\includegraphics[scale=0.37]{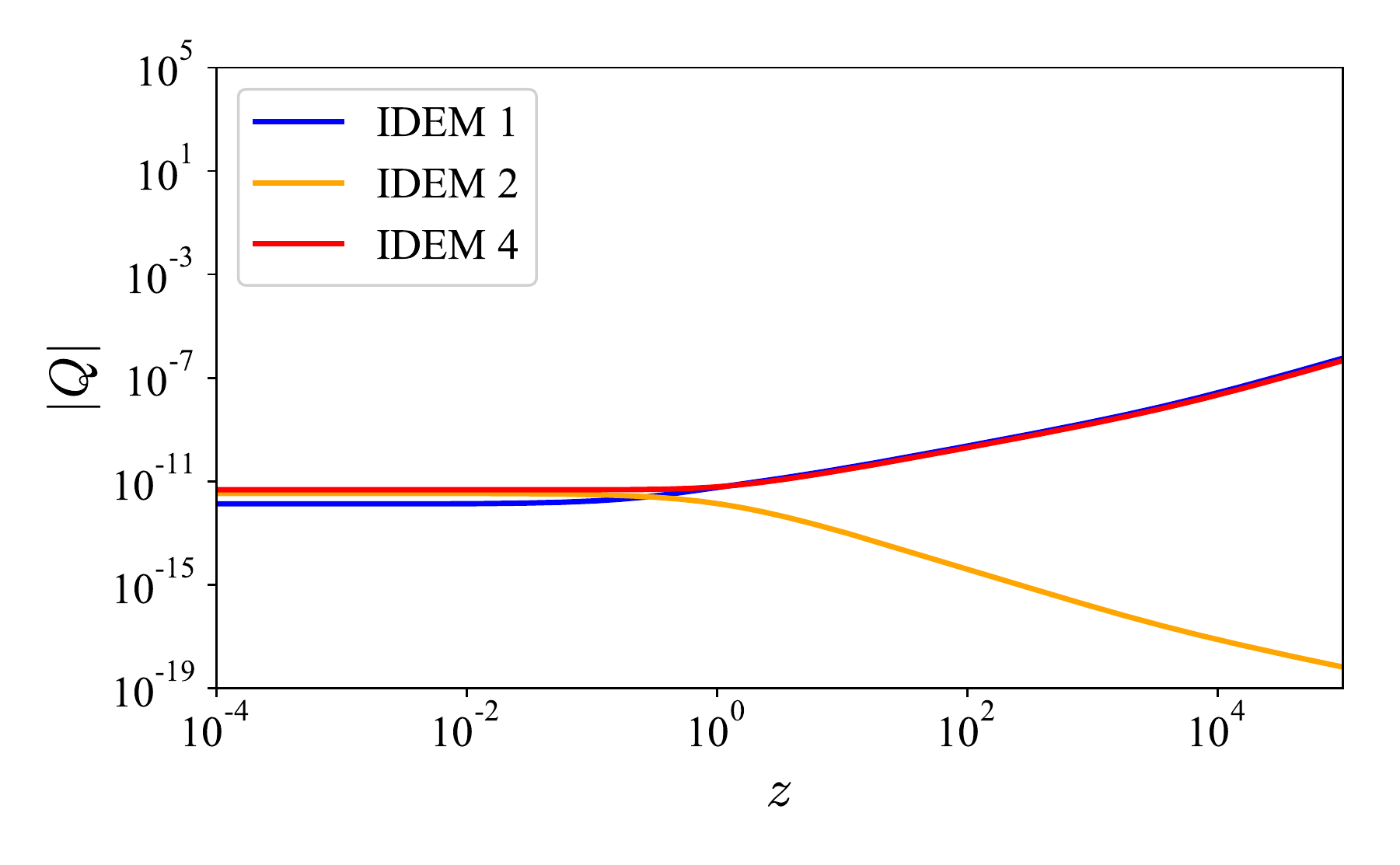}
\caption{Interaction parameter for all the 5 models. Left panel: Negative interaction parameter ($\gamma = -0.2$). Right panel: Positive interaction parameter ($\gamma=+0.2$).}
\label{idem}
\end{figure}

Our analysis reveals that the five investigated models can be divided into two groups.
The first group, comprised by IDEM 1, IDEM 2 and IDEM 4, is characterized by
interactions which become dynamically relevant only recently, i.e., close to
the present time. Technically, this is related to the proportionality of the interaction term
to (a power of) the DE density. The cosmological dynamics at high redshift
is almost unaffected and coincides with that of the standard model.
The second group is made up of IDEM 3 and IDEM 5. Here, the interaction term is proportional to
(a power of) the CDM energy density which means it is relevant already at early times.
Fig.~\ref{idem} shows the temporal evolution for all five models with $\gamma=-0.2$ (left panel) and $\gamma=+0.2$ (right panel).
At late times all models behave similarly, but at early times IDEM 3 and IDEM 5 differ strongly from the other models.
Fig.~\ref{idem} demonstrates that, even though $|Q|$ values of IDEM 1 and IDEM 4 are growing at high redshift, at $z\approx 10^{4}$ the $|Q|$ values for the models IDEM 3 and IDEM 5 are about eight orders of magnitude larger. In turn, IDEM 2 presents much lower values of $|Q|$ compared to the other models at recombination era.

\section{Statistical analysis}
\label{analysis}

In this section we present a statistical analysis for the IDEMs presented in the previous sections. The statistical analysis was performed through a suitable modification of the Boltzmann code CLASS \cite{Blas:2011rf} and the MCMC statistical code MontePython \cite{Audren:2012wb,Brinckmann:2018cvx}.

\subsection{Observational data}

In order to understand how each of the data sets constrains an interaction in the dark sector, the statistical analysis is performed gradually. At first we use geometrical tests related only to the recent expansion history of the Universe: Type-Ia Supernovae (SNe Ia), the present value of the Hubble rate ($H_{0}$) and Cosmic Chronometers (CC). Thereafter, we add Baryonic Acustic Oscilations (BAO) data which, while representing a geometrical test, are related to the primordial photon-baryon fluid. Finally, we constrain the models using the Planck TT data. We start by introducing the data sets used in the statistical analysis and their respective likelihoods.

\paragraph{\textbf{Type-Ia Supernovae (SNe Ia):}} The first data set used to perform statistical analysis is the SNe Ia data. Historically, the SNe Ia were
of great importance for cosmology, having been the key observation of the accelerated expansion observed currently \cite{Riess:1998cb,Perlmutter:1998np}. In
this work, we use the complete set of the “Joint Light-curve Analysis” (JLA) sample \cite{Betoule:2014frx}, which contains 740 data points from $z=0.01$
until $z=1.30$ \footnote{The data is available in \url{http://supernovae.in2p3.fr/sdss_snls_jla/jla_likelihood_v6.tgz}.}. The observable quantity in this case
is the is the distance modulus
\begin{equation}
\mu^{obs.} = m_{B}^{*}+\alpha\ x_{1}-\beta\ c-\mathcal{M}_{B}\,,
\label{muobs}
\end{equation}
where $m_{B}^{*}$ is the $B$-band peak magnitude measured in the rest-frame, $x_{1}$ is the time stretching of the light curve at maximum  brightness, $c$ is the color of the SN at maximum  brightness and $\mathcal{M}$ is related to the absolute $B$-band magnitude. The parameter $\mathcal{M}_{B}$ depends on the host stellar mass,
\begin{equation}
\mathcal{M}_{B}=\left\{\begin{array}{ll}M &\quad\mbox{if}\quad M_{stellar}\le 0,\\
M+\Delta_{M} &\quad\mbox{otherwise.}\end{array}\right.
\end{equation}
On the other hand, from a theoretical point of view, the distance modulus, in units of Mpc, can be obtained as
\begin{equation}
\mu^{th.}=5\log\left[\left(z+1\right)\int_{0}^{z}\dfrac{dz^{\prime}}{E\left(z^{\prime}\right)}\right]+25\,,
\label{muth}
\end{equation}
where the term in square brackets is the luminosity distance. The statistical analysis is then performed using the equations (\ref{muobs}) and (\ref{muth}) to write the likelihood function,
\begin{equation}
2\ln\left(\mathcal{L}_{SNe}\right)=\Delta\vec{\mu}^{T}\,\mathcal{C}^{-1}_{JLA}\,\Delta\vec{\mu}\,,
\end{equation}
where $\Delta\vec{\mu}$ is a vector whose components are $\mu^{obs.}_{i}-\mu^{th.}\left(z_{i}\right)$ and $\mathcal{C}_{JLA}$ is the covariance matrix of the JLA data, which is given by the sum of a statistical part and a systematic part ($\mathcal{C}_{JLA}=\mathcal{C}_{stat.}+\mathcal{C}_{sys.}$).

\paragraph{\textbf{Current value of the Hubble rate ($H_{0}$):}} As the second observable quantity, we use the recent model independent measure of the local value of the Hubble parameter from \cite{Riess:2016jrr}. Since SNe Ia can not constrain simultaneously the parameters $M$ and $H_{0}$,
we consider a combination of SNe Ia with this data point of $H_{0}$. The likelihood function in this case is
\begin{equation}
2\ln\left(\mathcal{L}_{H_{0}}\right)=\left(\dfrac{H_{0}-73.24}{1.74}\right)^{2}\,.
\end{equation}

\paragraph{\textbf{Cosmic Chronometers (CC):}} The third data set refers to the so-called cosmic chronometers. These are also model-independent data which are obtained from measures of differential ages of old galaxies that evolve passively at different times (different values of redshift, from $z=0.07$ until $z=1.75$). Combining these measures with the known redshift of the galaxy, one obtains the Hubble rate at the time. In this work we use the 31 data points presented in table \ref{tablecc}.
\begin{table}[h]
\centering
\begin{tiny}
\begin{tabular}{llll|llll|llll}
\hline\hline
$z\quad$    & $H(z)\quad$ & $\sigma_{H(z)}\quad$ & Ref.$\quad$& $z\quad$   & $H(z)\quad$ & $\sigma_{H(z)}\quad$ & Ref.$\quad$ & $z\quad$   & $H(z)\quad$ & $\sigma_{H(z)}\quad$ & Ref.$\quad$\\
\hline
0.07     & 69.0 & 19.6  & \cite{Zhang:2012mp}  & 0.4 & 95.0 & 17.0 & \cite{Simon:2004tf} & 0.875 & 125.0 & 17.0 & \cite{Moresco:2012jh}\\
0.09     & 69.0 & 12.0  & \cite{Simon:2004tf}    & 0.4004 & 77.0 & 10.2 & \cite{Moresco:2016mzx} & 0.88 & 90.0 & 40.0 & \cite{Stern:2009ep}\\
0.12      & 68.6 & 26.2 & \cite{Zhang:2012mp}  & 0.4247 & 87.1 & 11.2 & \cite{Moresco:2016mzx} & 0.9 & 117.0 & 23.0 & \cite{Simon:2004tf}    \\
0.17      & 83.0 & 8.0   & \cite{Simon:2004tf}     & 0.4497 & 92.8 & 12.9 & \cite{Moresco:2016mzx} & 1.037 & 154.0 & 20.0 & \cite{Moresco:2012jh}\\
0.179   & 75.0  & 4.0   & \cite{Moresco:2012jh} & 0.47 & 89.0 & 49.6 & \cite{Ratsimbazafy:2017vga} & 1.3 & 168.0 & 17.0 & \cite{Simon:2004tf}\\
0.199   & 75.0  & 5.0   & \cite{Moresco:2012jh} & 0.4783 & 80.9 & 9.0 & \cite{Moresco:2016mzx} & 1.363 & 160.0 & 33.6 & \cite{Moresco:2015cya}\\
0.2       & 72.9  & 29.6 & \cite{Zhang:2012mp}   & 0.48 & 97.0 & 62.0 & \cite{Stern:2009ep} & 1.43 & 177.0 & 18.0 & \cite{Simon:2004tf}\\
0.27     & 77.0  & 14.0 & \cite{Simon:2004tf}      & 0.593 & 104.0 & 13.0 & \cite{Moresco:2012jh} & 1.53 & 140.0 & 14.0 & \cite{Simon:2004tf}\\
0.28     & 88.8 & 36.6 & \cite{Zhang:2012mp}  & 0.68 & 92.0 & 8.0 & \cite{Moresco:2012jh} & 1.75 & 202.0 & 40.0 & \cite{Simon:2004tf}\\
0.352   & 83.0 & 14.0 & \cite{Moresco:2012jh} & 0.781 & 105.0 & 12.0 & \cite{Moresco:2012jh} & 1.965 & 186.0 & 50.4 & \cite{Moresco:2015cya}\\
0.3802 & 83.0 & 13.5 & \cite{Moresco:2016mzx}  & & & & & & & & \\
\hline\hline
\end{tabular}
\end{tiny}
\caption{Cosmic chronometers data.}
\label{tablecc}
\end{table}

Since all data points are independent, the likelihood function is given by,
\begin{equation}
2\ln\left(\mathcal{L}_{H\left(z\right)}\right)=\sum_{i=1}^{31}\left[\dfrac{H^{(i)}-H\left(z^{(i)}\right)}{\sigma^{(i)}_{H\left(z\right)}}\right]^{2}\,.
\end{equation}

\paragraph{\textbf{Baryonic Acustic Oscilations (BAO):}} The fourth data set used comes from the analysis of the baryonic acoustic oscilations.
Even if the BAO have a perturbative nature, they produce an imprint on the galaxy distribution that can be measured using background quantities.
The relevant physical quantities for the BAO data are the sound horizon at the drag time, the angular distance and the
spherically-averaged distance, which are given respectively by,
\begin{eqnarray}
r_{s}\equiv\int_{0}^{a_{drag}}\dfrac{c_{s}\left(a\right)}{Ha^{2}}\ da\,, \label{rs}\\
D_{A}\left(z\right)=\dfrac{1}{1+z}\int_{0}^{z}\dfrac{d\tilde{z}}{H\left(\tilde{z}\right)}\,, \\
D_{V}\left(z\right)=\left[\left(1+z\right)^{2}D_{A}^{2}\left(z\right)\dfrac{z}{H\left(z\right)}\right]^{1/3}\,,
\end{eqnarray}
where, in equation (\ref{rs}), $c_{s}$ corresponds to the sound speed in the primordial photon-baryon plasma. Table
\ref{tablebao} shows all the data used in this work with the respective surveys from where the data was obtained.

\begin{table}[h!]
\centering
\begin{tiny}
\begin{tabular}{c|c|c|c|c|c|c}\hline\hline
Catalog                    & $z$                 & BAO variable & BAO measurement                                 & $\sigma_{BAO}$ & $r_{s}^{fid}$       & Ref.                        \\ \hline
6dFGS                      & $0.106$                   & $\qquad \frac{r_{s}}{D_{V}}\qquad$                             & $0.327$   & $0.015$              &            *               & \cite{Beutler:2011hx}    \\ \hline
SDSS DR7 MGS               & $0.15$                    & $\qquad D_{V}\frac{r_{s}^{fid}}{r_{s}}\qquad$                  & $4.47$    & $0.16$               & $148.69$                  & \cite{Ross:2014qpa}        \\ \hline
BOSS-LOWZ                  & $0.32$                    & $\qquad D_{V}\frac{r_{s}^{fid}}{r_{s}}\qquad$                  & $8.47$    & $0.17$               & $149.28$                  & \cite{Anderson:2013zyy}    \\ \hline
\multirow{6}{*}{BOSS-DR12} & \multirow{2}{*}{$0.38$} & $\qquad D_{A}\left(1+z\right)\frac{r_{s}^{fid}}{r_{s}}\qquad$  & $1512.39$ & $25.00$              & \multirow{6}{*}{$147.78$} & \multirow{6}{*}{\cite{Alam:2016hwk}} \\
                           &                           & $\qquad H\frac{r_{s}}{r_{s}^{fid}}\qquad$                      & $81.2087$ & $2.3683$             &                           &                             \\ \cline{2-5}
                           & \multirow{2}{*}{$0.51$} & $\qquad D_{A}\left(1+z\right)\frac{r_{s}^{fid}}{r_{s}}\qquad$  & $1975.22$ & $30.10$              &                           &                             \\
                           &                           & $\qquad H\frac{r_{s}}{r_{s}^{fid}}\qquad$                      & $90.9029$ & $2.3288$             &                           &                             \\\cline{2-5}
                           & \multirow{2}{*}{$0.61$} & $\qquad D_{A}\left(1+z\right)\frac{r_{s}^{fid}}{r_{s}}\qquad$  & $2306.68$ & $37.08$              &                           &                             \\
                           &                           & $\qquad H\frac{r_{s}^{fid}}{r_{s}}\qquad$                      & $98.9647$ & $2.5019$             &                           &                             \\ \hline
\multirow{3}{*}{WiggleZ}   & $0.44$                    & \multirow{3}{*}{$\qquad D_{V}\frac{r_{s}^{fid}}{r_{s}}\qquad$} & $1716$    & $83$                 & \multirow{3}{*}{$148.6$}  & \multirow{3}{*}{\cite{Kazin:2014qga}}  \\
                           & $0.60$                    &                                                                & $2221$    & $101$                &                           &                             \\
                           & $0.73$                    &                                                                & $2516$    & $86$                 &                           &                             \\ \hline
BOSS-CMASS                 & $0.57$                    & $\qquad D_{V}\frac{r_{s}^{fid}}{r_{s}}\qquad$                  & $13.77$   & $0.13$               & $149.28$                  & \cite{Anderson:2013zyy}    \\ \hline\hline
\end{tabular}
\end{tiny}
\caption{BAO data.}
\label{tablebao}
\end{table}

In general, the BAO likelihood takes the following form,
\begin{equation}
2\ln\left(\mathcal{L}_{BAO}\right)=\Delta\vec{V}^{T}\,\mathcal{C}^{-1}_{BAO}\,\Delta\vec{V}\,.
\end{equation}
In the above equation, $\Delta\vec{V}$ is a vector whose components are given by $V^{obs.}_{i}-V^{th.}\left(z_{i}\right)$,
where $V$ corresponds to the BAO variables in the third column of table \ref{tablebao}, and $\mathcal{C}_{BAO}$ is the
covariance matrix of the data. In this case, only the data from WiggleZ and BOSS-DR12 are correlated, and their respective covariance matrices are
\begin{eqnarray}
\mathcal{C}_{WiggleZ}^{-1}=10^{-4}\left( \begin{array}{ccc}
2.17898878 & -1.11633321 & 0.46982851 \\
 & 1.70712004 & -0.71847155 \\
 &  & 1.65283175 \end{array} \right)
\\
\mathcal{C}_{BOSS-DR12}=10^{4}\left( \begin{array}{cccccc}
624.707 & 23.729 & 325.332 & 8.34963 & 157.386 & 3.57778 \\
 & 5.60873 & 11.6429 & 2.33996 & 6.39263 & 0.968056 \\
 &  & 905.777 & 29.3392 & 515.271 & 14.1013 \\
 &  &  & 5.42327 & 16.1422 & 2.85334 \\
 &  &  &  & 1375.12 & 40.4327 \\
 &  &  &  &  & 6.25936 \end{array} \right)
\end{eqnarray}

Recently, a statistical analysis using data from angular BAO  was performed in \cite{Cid:2018ugy}.

\paragraph{\textbf{Planck TT:}} The last data set used to constrain the interacting models is the Planck measurements of the CMB temperature anisotropy. As
it is well-know, the CMB data is able to provide a strong constraint on the parameter $\Omega_{c0}$, then, since the interaction parameter $\gamma$
affects the CDM dynamics (as well as the DE dynamics), it is also expected that the CMB data can strongly constrain the interacting models.
In this work, we use the Commander and Plik codes \footnote{Both, likelihood code and data, are available in \url{http://pla.esac.esa.int/pla/\#cosmology}.},
respectively, for the low $l$ analysis ($l<30$), and for the high $l$ analysis ($l\geq30$) \cite{Aghanim:2015xee}.

\subsection{Results}

\begin{figure}[h!]
\includegraphics[scale=0.7]{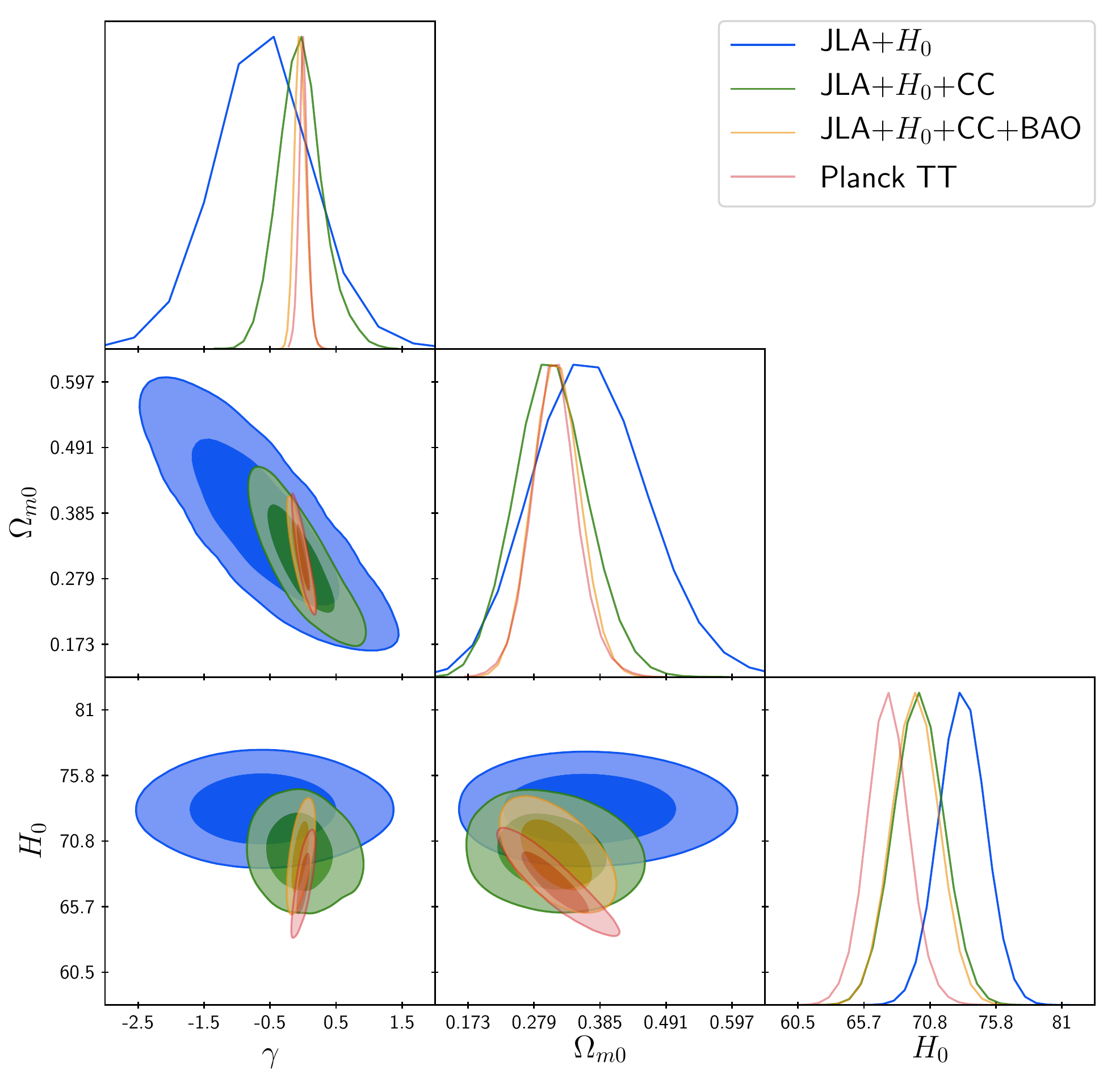}
\caption{Statistical analysis IDEM 1.}
\label{idem1_triangle}
\end{figure}

\begin{figure}[h!]
\includegraphics[scale=0.7]{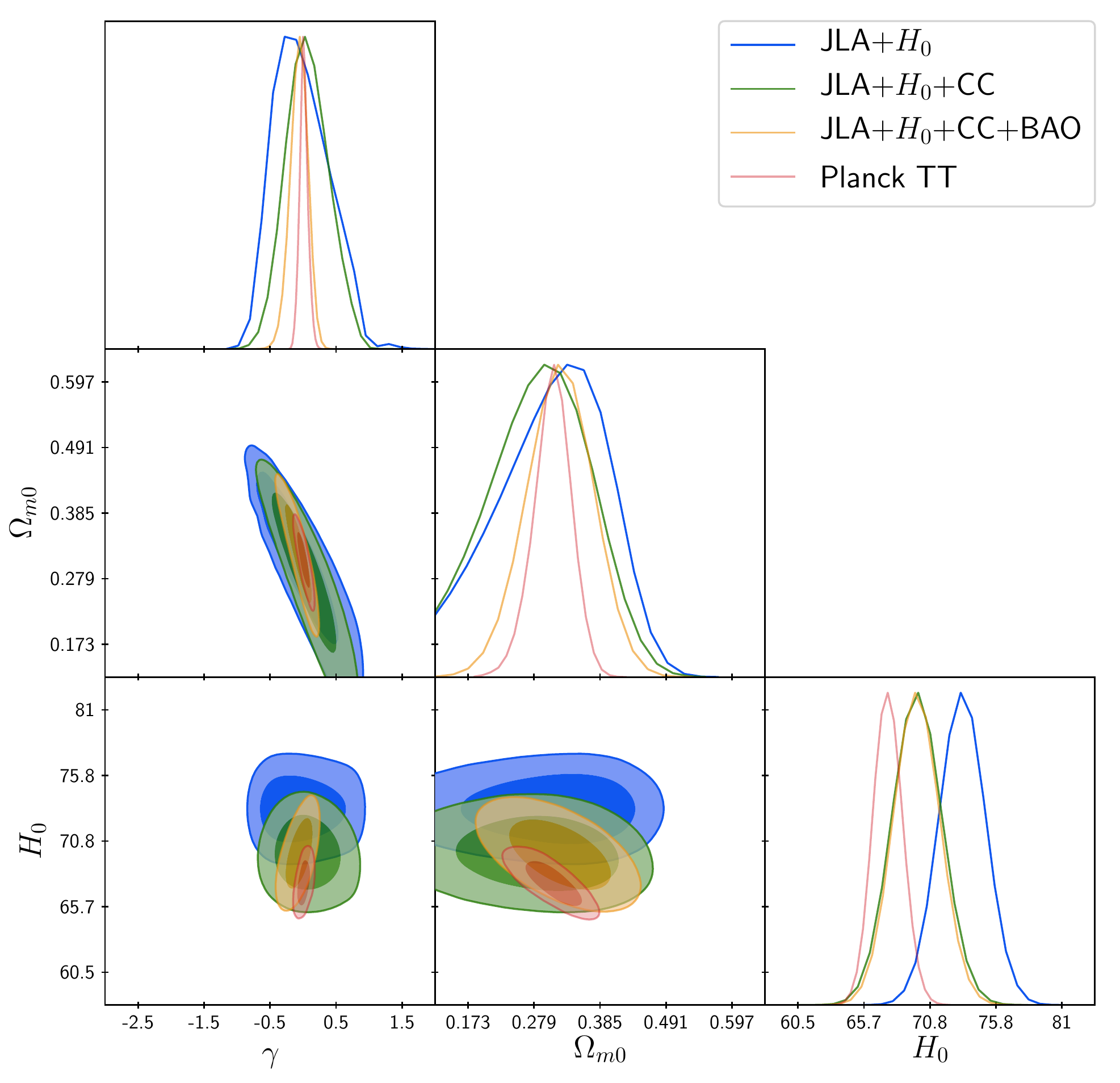}
\caption{Statistical analysis IDEM 2.}
\label{idem2_triangle}
\end{figure}

\begin{figure}[h!]
\includegraphics[scale=0.7]{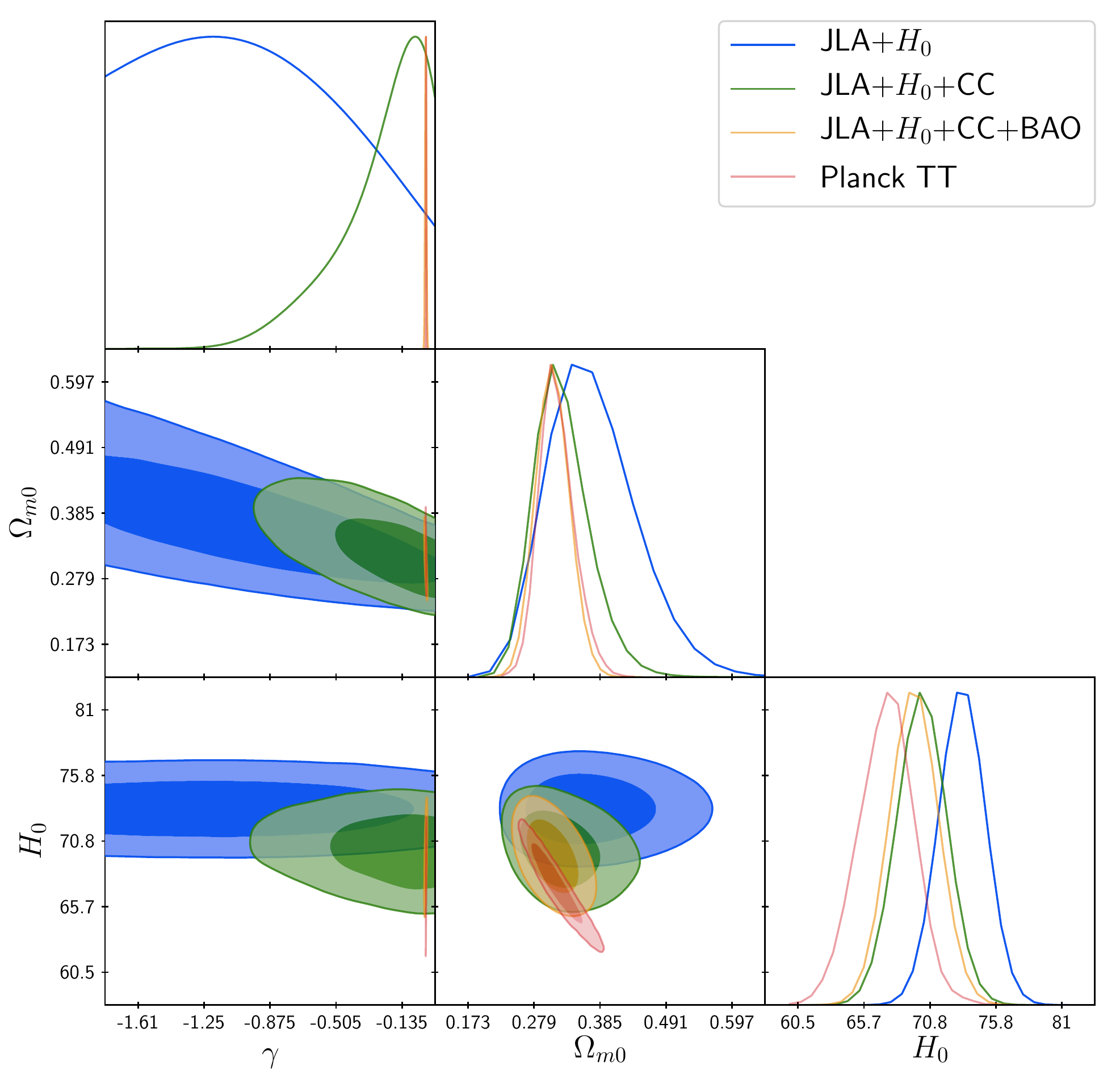}
\caption{Statistical analysis IDEM 3.}
\label{idem3_triangle}
\end{figure}

\begin{figure}[h!]
\includegraphics[scale=0.7]{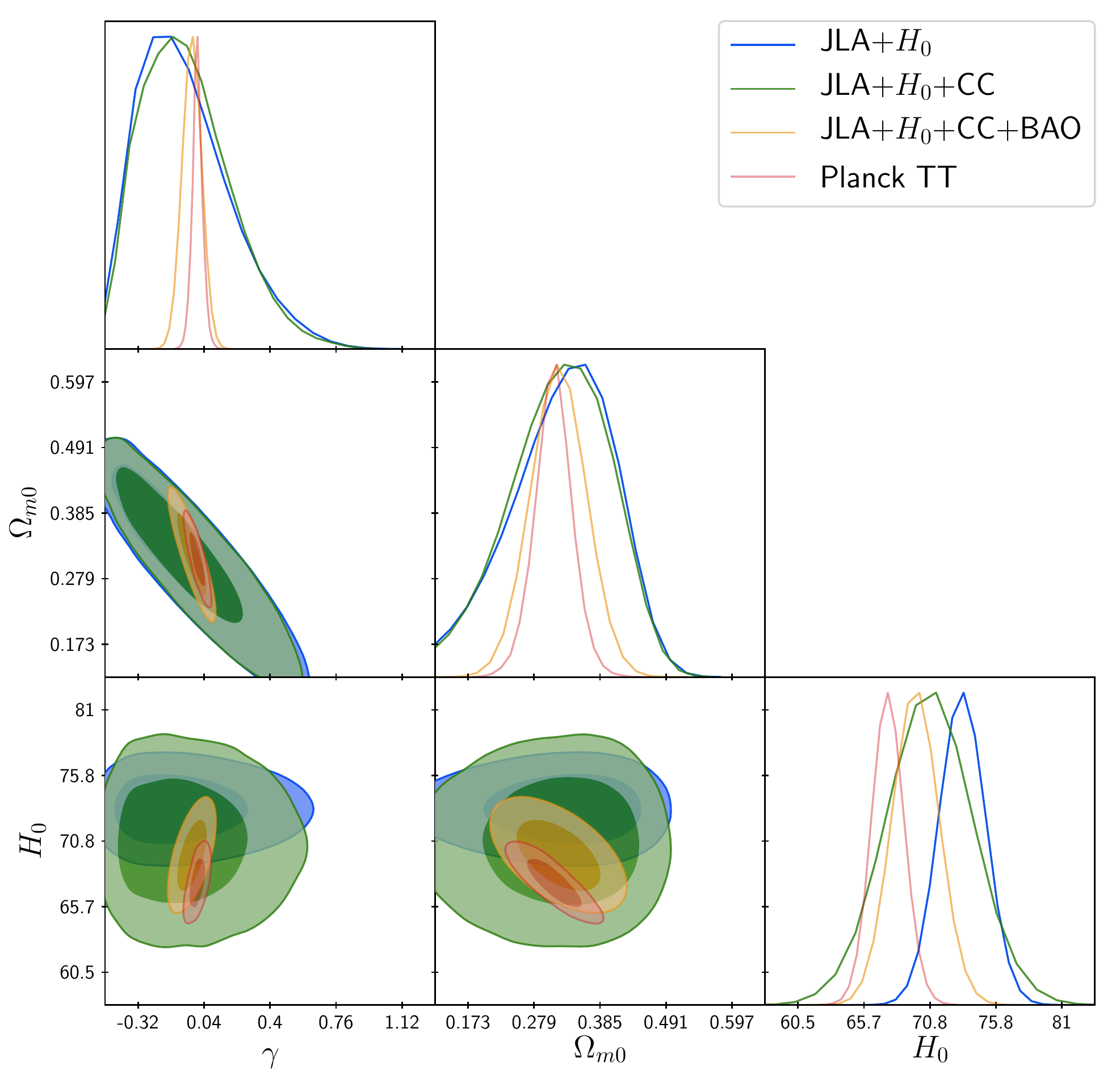}
\caption{Statistical analysis IDEM 4.}
\label{idem4_triangle}
\end{figure}

\begin{figure}[h!]
\includegraphics[scale=0.7]{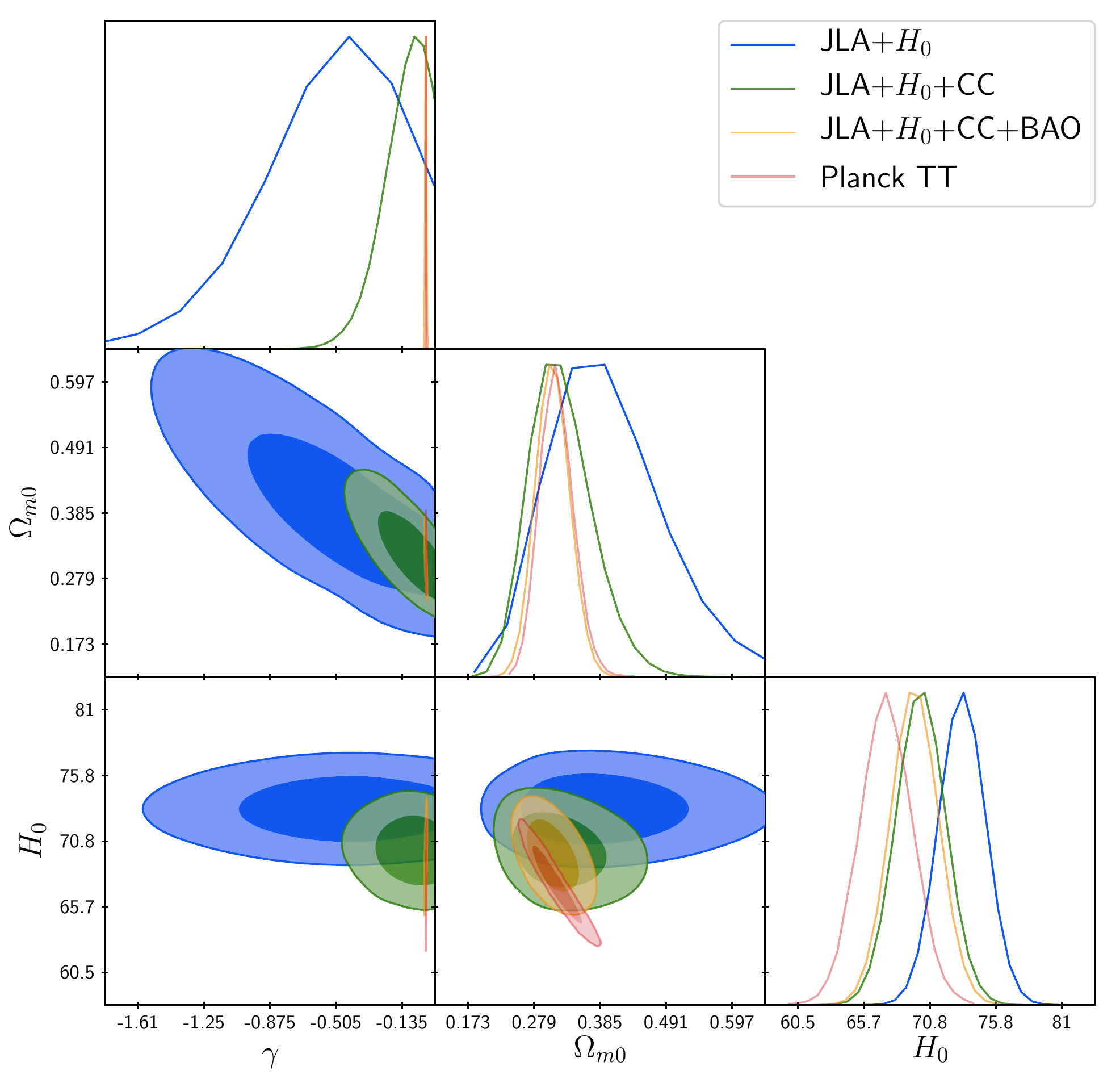}
\caption{Statistical analysis IDEM 5.}
\label{idem5_triangle}
\end{figure}

Our results are summarized in tables \ref{tableidem} and \ref{tableidemcmb}, and figures \ref{idem1_triangle}, \ref{idem2_triangle}, \ref{idem3_triangle}, \ref{idem4_triangle}, and \ref{idem5_triangle}.
In table \ref{tableidem} we list the values for $H_{0}$, $\Omega_{m0}$ and $\gamma$ where $\Omega_{m0}$ is the total matter density parameter, defined by the sum of CDM and baryon contributions. For the background tests the baryon density parameter was fixed by the results from nucleosynthesis \cite{Pettini:2012ph}, in the Planck TT analysis, however,  $\Omega_{b0}$ is a free parameter. Table \ref{tableidemcmb} shows the results for $\Omega_{b0}h^{2}$, $\Omega_{c0}h^{2}$, the actual angular scale of the sound horizon at decoupling $100\theta_{s}$, the spectral tilt $n_{s}$ and the reionization parameter $\tau_{reio }$ if only the Planck TT data are used.
We mention that  for all tests, the Gelman-Rubin convergence parameter satisfies the condition $\hat{R}-1<0.01$ \cite{Gelman:1992zz}.

\begin{table}[h!]
\centering
\begin{tiny}
\begin{tabular}{l|l|c|c|c|c}
\hline \hline
Model & Data & $H_{0}$ & $\Omega_{m0}$ & $\gamma$ & $\quad\chi^2_{min}\quad$ \\ \hline
\multirow{ 4}{*}{IDEM1} & SNe Ia+$H_{0}$        & $73.37_{-3.61}^{+3,63}$ & $0.354_{-0.162}^{+0.109}$ & $-0.53_{-0.91}^{+1.02}$ & $682.14$ \\
                        & SNe Ia+$H_{0}$+CC     & $70.78_{-3.61}^{+3.62}$ & $0.307_{-0.122}^{+0.108}$ & $-0.07_{-0.74}^{+0.58}$ & $695.45$ \\
                        & SNe Ia+$H_{0}$+CC+BAO & $69.44_{-3.61}^{+3.62}$ & $0.321_{-0.078}^{+0.072}$ & $-0.06_{-0.18}^{+0.16}$ & $697.77$ \\
                        & Planck TT             & $68.13_{-2.96}^{+2.86}$ & $0.3143_{-0.0685}^{+0.0636}$ & $-0.010_{-0.140}^{+0.108}$ & $11261.4$ \\
\hline
\multirow{ 4}{*}{IDEM2} & SNe Ia+$H_{0}$        & $73.29_{-3.60}^{+3.60}$ & $0.371_{-0.136}^{+0.189}$ & $-0.40_{-0.94}^{+0.58}$ & $682.24$ \\
                        & SNe Ia+$H_{0}$+CC     & $69.75_{-3.60}^{+3.60}$ & $0.307_{-0.148}^{+0.186}$ & $-0.04_{-0.70}^{+0.60}$ & $695.46$ \\
                        & SNe Ia+$H_{0}$+CC+BAO & $69.72_{-3.60}^{+3.60}$ & $0.326_{-0.104}^{+0.110}$ & $-0.08_{-0.24}^{+0.28}$ & $697.81$ \\
                        & Planck TT             & $68.00_{-2.47}^{+2.28}$ & $0.3054_{-0.050}^{+0.054}$ & $-0.0024_{-0.105}^{+0.104}$ & $11262.2$ \\
\hline
\multirow{ 4}{*}{IDEM3} & SNe Ia+$H_{0}$        & $73.21_{-3.60}^{+3.60}$ & $0.370_{-0.136}^{+0.196}$ & $-0.23_{-1.31}^{+0.23}$ & $682.22$ \\
                        & SNe Ia+$H_{0}$+CC     & $70.70_{-3.60}^{+3.60}$ & $0.381_{-0.138}^{+0.194}$ & $-0.27_{-0.31}^{+0.46}$ & $682.23$ \\
                        & SNe Ia+$H_{0}$+CC+BAO & $69.64_{-3.60}^{+3.60}$ & $0.320_{-0.088}^{+0.088}$ & $-0.038_{-0.22}^{+0.24}$ & $697.74$ \\
                        & Planck TT             & $67.35_{-2.00}^{+2.41}$ & $0.3157_{-0.0641}^{+0.0465}$ & $\left(1.36_{-8.47}^{+9.62}\right)\times10^{-06}$ & $11262.0$ \\
\hline
\multirow{ 4}{*}{IDEM4} & SNe Ia+$H_{0}$        & $73.21_{-3.61}^{+3.60}$ & $0.379_{-0.239}^{+0.097}$ & $-0.26_{-0.24}^{+0.70}$ & $682.22$ \\
                        & SNe Ia+$H_{0}$+CC     & $70.70_{-4.82}^{+4.81}$ & $0.379_{-0.238}^{+0.097}$ & $-0.27_{-0.24}^{+0.68}$ & $682.24$ \\
                        & SNe Ia+$H_{0}$+CC+BAO & $69.64_{-3.61}^{+3.60}$ & $0.320_{-0.089}^{+0.087}$ & $-0.037_{-0.20}^{+0.22}$ & $697.75$ \\
                        & Planck TT             & $67.51_{-2.66}^{+2.47}$ & $0.309_{-0.058}^{+0.058}$ & $-0.0052_{-0.098}^{+0.102}$ & $11262.1$ \\
\hline
\multirow{ 4}{*}{IDEM5} & SNe Ia+$H_{0}$        & $73.24_{-3.61}^{+3.60}$ & $0.362_{-0.220}^{+0.134}$ & $-0.36_{-0.60}^{+0.36}$ & $682.11$ \\
                        & SNe Ia+$H_{0}$+CC     & $70.05_{-3.61}^{+3.60}$ & $0.307_{-0.108}^{+0.080}$ & $-0.092_{-0.22}^{+0.10}$ & $695.45$ \\
                        & SNe Ia+$H_{0}$+CC+BAO & $69.56_{-3.61}^{+3.60}$ & $0.309_{-0.029}^{+0.026}$ & $-0.0019_{-0.0072}^{+0.0070}$ & $697.83$ \\
                        & Planck TT             & $67.36_{-2.53}^{+2.81}$ & $0.3154_{-0.030}^{+0.024}$ & $\left(-9.73_{-8.37}^{+8.82}\right)\times10^{-05}$ & $11262.2$ \\
\hline \hline
\end{tabular}
\end{tiny}
\caption{Result of the statistical analysis with $2\sigma$ CL for all IDEMs.}
\label{tableidem}
\end{table}

\begin{table}[]
\centering
\begin{tiny}
\begin{tabular}{l|c|c|c|c|c}
\hline\hline
Parameters  & IDEM 1 & IDEM 2 & IDEM 3 & IDEM 4 & IDEM 5 \\ \hline
$\Omega_{b0}h^{2}$ & $0.02217_{-0.00048}^{+0.00046}$ & $0.02235_{-0.00048}^{+0.00048}$ & $0.02220_{-0.00052}^{+0.00052}$ & $0.022320_{-0.00048}^{+0.00046}$ & $0.02228_{-0.052}^{+0.050}$ \\
$\Omega_{c0}h^{2}$ & $0.1121_{-0.0196}^{+0.0200}$ & $0.1299_{-0.0166}^{+0.0240}$ & $0.1198_{-0.0078}^{+0.0068}$ & $0.1190_{-0.0158}^{+0.0190}$ & $0.1248_{-0.0076}^{+0.0070}$ \\
$100\theta_{s}$ & $1.042_{-0.00092}^{+0.00092}$ & $1.042_{-0.00092}^{+0.00092}$ & $1.042_{-0.00090}^{+0.00090}$ & $1.042_{-0.00090}^{+0.00092}$ & $1.042_{-0.00082}^{+0.00088}$ \\
$ln\left(10^{10}A_{s}\right)$ & $3.073_{-0.076}^{+0.076}$ & $3.086_{-0.076}^{+0.074}$ & $3.069_{-0.074}^{+0.068}$ & $3.095_{-0.076}^{+0.074}$ & $3.073_{-0.070}^{+0.072}$ \\
$n_{s}$ & $0.9751_{-0.0132}^{+0.0128}$ & $0.9637_{-0.0130}^{+0.0126}$ & $0.9627_{-0.0138}^{+0.0132}$ & $0.9673_{-0.0065}^{+0.0063}$ & $0.9571_{-0.0065}^{+0.0067}$ \\
$\tau_{reio }$ & $0.07477_{-0.020}^{+0.019}$ & $0.07585_{-0.020}^{+0.019}$ & $0.06931_{-0.019}^{+0.018}$ & $0.08166_{-0.020}^{+0.019}$ & $0.06573_{-0.019}^{+0.019}$
\\ \hline\hline
\end{tabular}
\end{tiny}
\caption{Result of the statistical analysis with $2\sigma$ CL for all IDEMs using only Planck TT data.}
\label{tableidemcmb}
\end{table}

\section{Discussion and Conclusions}
\label{conclusions}

We investigated five types of dark-sector interactions for which the ratio $r$ of the
energy densities of CDM and DE is a function of the scale factor only.
These models are examples of a general class for which a unified description in terms of a function
$f\left(r\right)$ is possible.
Requiring that early DE be not negative provides us with constraints on the interaction parameter $\gamma$.
For models IDEM 2 and IDEM 4 we have a negative lower bound on $\gamma$,
while for IDEM 3 and IDEM 5 there is an upper  limit $\gamma = 0$, which
means that only matter creation is allowed in these models.

Models IDEM 3 and IDEM 5 are particularly sensitive to the interaction, even a small
value of the interaction parameter  can drastically affect the background dynamics.
As shown in Fig.\ref{idem}  the interaction for IDEM 3 and IDEM 5 at high redshift is stronger than
that for the other models. This indicates a rather high rate of matter creation at an early epoch.
Since we don't expect a present amount of CDM much bigger than the standard-model value,
such interacting models necessarily have a rather low CDM fraction in the past.
As Figs.~\ref{idem3_rho} and \ref{idem5_rho} show, these models, upon assuming  the standard value of $\Omega_{c0}$, predict a baryon-dominated era,
which does not seem to be compatible with the standard description of the Universe
before and through the recombination era.

The late-time observational data from $H_{0}$ and SNIa leave room for a matter creation scenario
($\gamma < 0$). The data from BAO, or above all, from Planck TT, however, constrain the interaction strongly
to values very close to $\gamma = 0$ ($\Lambda$CDM model).  Models IDEM 3 and IDEM 5 are virtually discarded.
The remaining models allow for a small range of the interaction parameter ($\gamma \sim \pm 0.15$ at the 2$\sigma$ confidence level).

Since the $\Lambda$CDM model fits most observation extremely well, in particular the CMB data, from the outset,
interacting models are not expected to disagree substantially from this standard-model behavior.

Quite generally, our analysis demonstrates that an interaction in the dark sector is not excluded but the range
for the still admissible interaction parameter is very narrow, being always consistent in $1\sigma$ CL 
with the zero coupling case ($\gamma=0$). 
Moreover, for viable models the interaction has to become dynamically relevant only close to the present time.
An extended
analysis of these models (for example at non-linear level) may indicate that even this small interaction can lead to interesting results.

\appendix
\section{The SNe Ia (JLA) analysis}

\begin{figure}[h!]
\includegraphics[scale=0.6]{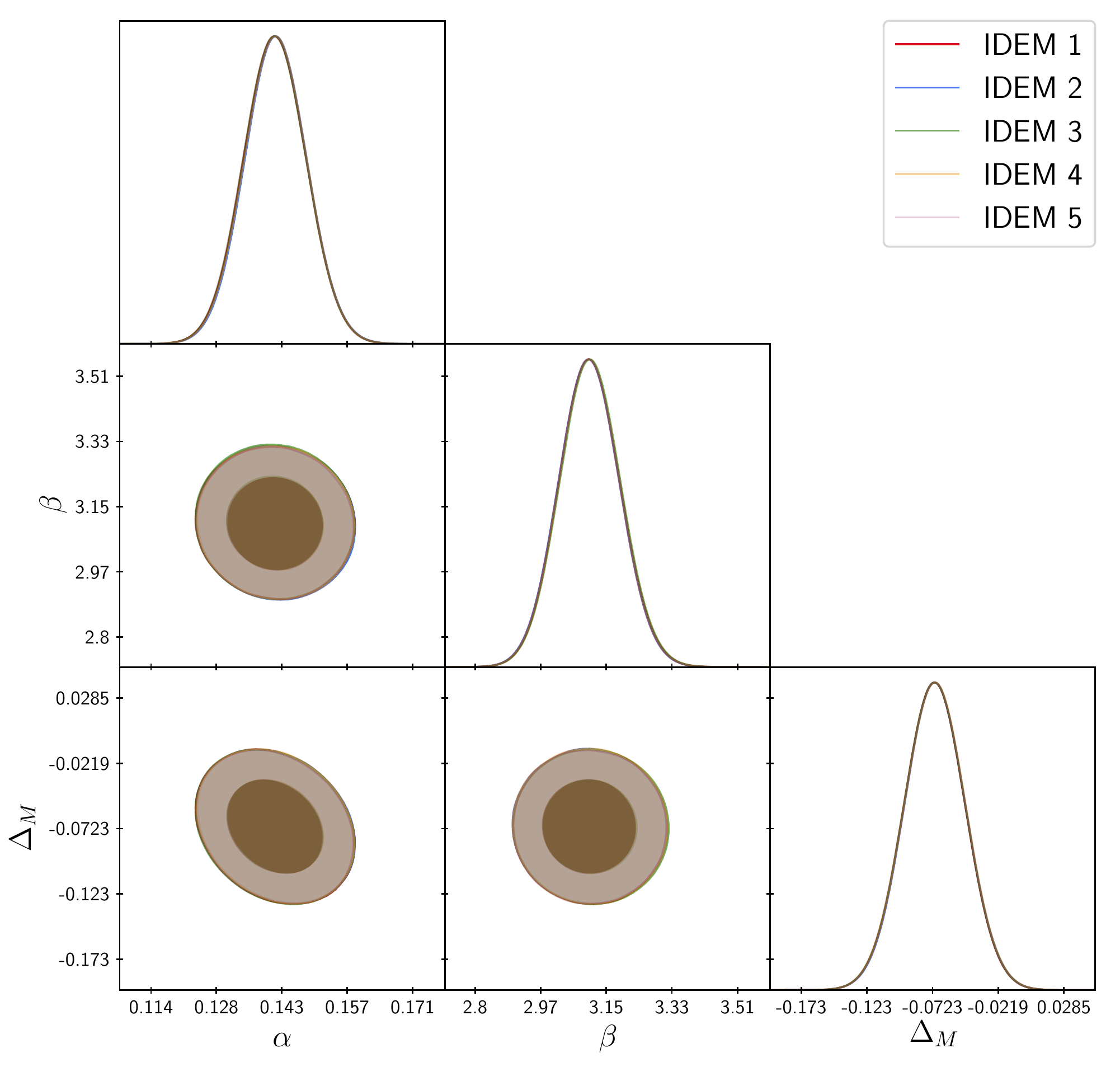}
\caption{Statistical analysis using the SNe Ia (JLA) data for the nuisance parameters $\alpha$, $\beta$ and $\Delta_{M}$.}
\label{nuisance}
\end{figure}

The authors of reference \cite{Betoule:2014frx} mention that the correlation between $\Omega_{m0}$ and the nuisance parameters $\alpha$, $\beta$ and $\Delta_{M}$ is small for the $\Lambda$CDM model. This fact suggest that for models with isotropic luminosity distance which are evolving smoothly with redshift, the binned JLA data can be a reasonable data set to constrain the cosmological parameters.  Now, in the context of the present paper,  the interaction parameter $\gamma$ affects the CDM dynamics (as well as the DE dynamics).
Therefore, it seems prudent to verify if there is a correlation between $\gamma$ and the nuisance parameters $\alpha$, $\beta$ and $\Delta_{\mbox{M}}$. In this appendix we present a statistical analysis for all the investigated models using only the SNe Ia (JLA) data. The result is shown in table \ref{tablejla} and in figure \ref{nuisance}.
In this analysis we marginalize numerically over the combination of the parameters $M$ and $H_{0}$. According our results, one can conclude that the nuisance parameters are almost unaffected by the interaction.

\begin{table}
\centering
\begin{tiny}
\begin{tabular}{l|c|c|c|c|c}
 \hline \hline
Model & $\Omega_{m0}$ & $\gamma$ & $\alpha$ & $\beta$ & $\Delta_{M }$ \\ \hline
$\Lambda$CDM & $0.295_{-0.068}^{+0.072}$ & $0$ & $0.1412_{-0.0133}^{+0.0137}$ & $3.098_{-0.156}^{+0.176}$ & $-0.0698_{-0.0478}^{+0.0473}$ \\
IDEM1 & $0.365_{-0.172}^{+0.188}$ & $-0.579_{-2.696}^{+1.505}$ & $0.1404_{-0.0130}^{+0.0143}$ & $3.099_{-0.158}^{+0.174}$ & $-0.0703_{-0.0479}^{+0.0472}$ \\
IDEM2 & $0.376_{-0.267}^{+0.086}$ & $-0.400_{-0.314}^{+1.192}$ & $0.1410_{-0.0132}^{+0.0139}$ & $3.100_{-0.160}^{+0.172}$ & $-0.0705_{-0.0473}^{+0.0476}$ \\
IDEM3 & $0.353_{-0.142}^{+0.382}$ & $-0.813_{-3.615}^{+0.955}$ & $0.1410_{-0.1355}^{+0.0137}$ & $3.106_{-0.165}^{+0.170}$ & $-0.0711_{-0.0476}^{+0.0477}$ \\
IDEM4 & $0.380_{-0.240}^{+0.095}$ & $-0.263_{-0.228}^{+0.709}$ & $0.1406_{-0.0129}^{+0.0142}$ & $3.101_{-0.162}^{+0.171}$ & $-0.0703_{-0.0480}^{+0.0932}$ \\
IDEM5 & $0.359_{-0.168}^{+0.135}$ & $-0.356_{-0.485}^{+0.942}$ & $0.1409_{-0.0131}^{+0.0140}$ & $3.103_{-0.163}^{+0.168}$ & $-0.0698_{-0.0540}^{+0.0410}$ \\
\hline \hline
 \end{tabular}
 \end{tiny}
\caption{Statistical analysis with $2\sigma$ CL using the SNe Ia (JLA) data for the nuisance parameters.}
\label{tablejla}
\end{table}

\ \\
\noindent
{\bf Acknowledgement:} We thank S.D.P. Vitenti for valuable instructions about numerical issues and useful discussions. Financial support by CNPq, CAPES and FAPES is gratefully acknowledged. The simulations were performed on resources provided by UNINETT Sigma2 – the National Infrastructure for High Performance Computing and Data Storage in Norway. This work has made use of the computing facilities of NPAD/UFRN, and of the Laboratory of Astroinformatics (IAG/USP, NAT/Unicsul), whose purchase was made possible by the Brazilian agency FAPESP (2009/54006-4) and the INCT-A. RvM acknowledges support from Federal Commission for Scholarships for Foreign Students for the Swiss Government Excellence Scholarship (ESKAS No. 2018.0443) for the academic year 2018–2019.




\end{document}